\documentclass[11pt,a4paper]{article}
\pdfoutput=1

\usepackage{bmpsize}
\usepackage{jheppub}
\usepackage{graphicx,bm,color,wrapfig,cancel}

\usepackage{amsmath,amssymb,bm}
\usepackage{float}
\usepackage{braket}
\usepackage{comment}
\usepackage{subfigure}
\usepackage[utf8]{inputenc}
\usepackage{appendix}
\usepackage{ulem}

\newcommand{\fulltoday}{\number\day\space \ifcase\month\or
    January\or February\or March\or April\or May\or June\or
    July\or August\or September\or October\or November\or December\fi
    \space\number\year}

 \makeatletter
    
    \@addtoreset{equation}{section}
  \makeatother

\allowdisplaybreaks[3]

\title{\boldmath
Dynamic scale anomalous transport in QCD with electromagnetic background}

\author[a]{Mamiya Kawaguchi,}
\author[b]{Shinya Matsuzaki,}
\author[a,c]{Xu-Guang Huang}

\affiliation[a]{Department of Physics and Center for Field Theory and Particle Physics, Fudan University, 220 Handan Road, 200433 Shanghai, China}

\affiliation[b]{Center for Theoretical Physics and College of Physics, Jilin University, Changchun, 130012, China}
\affiliation[c]{Key Laboratory of Nuclear Physics and Ion-beam Application (MOE), Fudan University, Shanghai 200433, China}

\emailAdd{kawaguchi@fudan.edu.cn}
\emailAdd{synya@jlu.edu.cn}
\emailAdd{huangxuguang@fudan.edu.cn}

\abstract{
We discuss phenomenological implications of the anomalous transport induced by the scale anomaly in QCD coupled to an electromagnetic (EM) field, based on a dilaton effective theory. The scale anomalous current emerges in a way perfectly analogous to the conformal transport current induced in a curved spacetime background, or the Nernst current in Dirac and Weyl semimetals -- both current forms are equivalent by a ``Weyl transformation". We focus on a spatially homogeneous system of QCD hadron phase, which is expected to be created after the QCD phase transition and thermalization. We find that the EM field can induce a dynamic oscillatory dilaton field which in turn induces the scale anomalous current. As the phenomenological applications, we evaluate the dilepton and diphoton productions induced from the dynamic scale anomalous current, and find that those productions include a characteristic peak structure related to the dynamic oscillatory dilaton, which could be tested in heavy ion collisions. We also briefly discuss the out-of-equilibrium particle production created by a nonadiabatic dilaton oscillation, which happens in a way of the so-called tachyonic preheating mechanism.
}

\begin{document}
\maketitle
\flushbottom

\section{Introduction}

The transport phenomena induced by chiral anomaly (dubbed chiral anomalous transports) have been extensively studied in heavy-ion collisions where both the strong electromagnetic (EM) fields (of the order of $m_\pi^2$ with $m_\pi$ pion mass) and the local parity odd domains with net chirality (due to, e.g., strong CP violating topological transition) are generated. The search of such chiral anomalous transports is one of the frontiers of current heavy-ion collision experiments and, if they are confirmed, it would significantly boost our understanding of the quantum chromodynamics (QCD) vacuum structure; see Refs. ~\cite{Kharzeev:2015znc,Hattori:2016emy,Zhao:2019hta,Liu:2020ymh} for reviews. 

Nowadays, the target field in which such chiral anomalous transports play some role has been extended with multiple aspects including  not only QCD and condensed matter systems, but also applications to other cosmologically and theoretically important issues, such as production of the baryon number asymmetry of universe~\cite{Joyce:1997uy,Bamba:2006km,Kamada:2016eeb,Kamada:2016cnb,Anber:2015yca,Jimenez:2017cdr,Domcke:2019mnd,Domcke:2019qmm}, inflationary scenarios with axion~\cite{Turner:1987vd,Garretson:1992vt,Anber:2006xt}, and a solution for the gauge-hierarchy problem~\cite{Hook:2016mqo,Choi:2016kke,Tangarife:2017vnd,Tangarife:2017rgl,Fonseca:2018xzp}. 
Thus, the anomalous transport physics has opened a vast  
ballpark in the theoretical particle physics.

Paying an attention to another candidate for possible emergence of anomalous transports in QCD,
one may notice the scale symmetry.
The scale symmetry of QCD is explicitly broken and anomalous due to quantum corrections, which
is dictated even at the one-loop perturbative beta function
in pure gluonic QCD (Yang-Mills).
Coupled to quarks, the QCD scale current externally gets other anomalous parts from the quark mass terms, and electroweak gauge corrections which quarks
perturbatively feel.
Among them, in particular, the external EM fields
should give a significant contribution to
the induced anomalous-scale transport, with relevance enough to leave
some phenomenological implications, in heavy ion collisions and/or early universe during the thermal history where strong EM fields exist as well.
In this perspective, the anomalous transport from the scale anomaly (will be dubbed scale anomalous transport hereafter) might have
provide an effect similar to the case with the chiral anomaly.

Such an EM-field-induced scale anomalous transport has recently been
discussed in the contexts different from QCD~\cite{Chernodub:2016lbo,Chernodub:2017bbd,Chernodub:2017jcp,Zheng:2019xeu}.
The electric current $j_\mu$ has been proposed to emerge in
quantum electrodynamics (QED) in a curved spacetime background :
\begin{eqnarray}
\langle j^\mu(x)\rangle=-\frac{2\beta_{}(e)}{e}F^{\mu\nu}(x)\partial_\nu\tau(x)
+\frac{2\beta_{}(e)}{e}\left(\partial_\nu F^{\nu\mu}(x)\right)\tau(x),
\label{ano_QEDscale}
\end{eqnarray}
where $\tau(x)$ represents the scale factor of the curved spacetime metric, $g_{\mu\nu}=e^{2\tau(x)} \eta_{\mu\nu}$ with $\eta_{\mu\nu}$ being the flat spacetime metric,
and
$\beta_{}(e)$ is the beta function of the EM coupling $e$.
Phenomenological applications have also been discussed
in condensed matter systems, such as Dirac and Weyl semimetals, that was called induced Nernst current~\cite{Chernodub:2017jcp}.






We note that although the current $j_\mu$ in Eq.~(\ref{ano_QEDscale}) completely relies on the existence of the scale factor $\tau(x)$ of the curved background spacetime which vanishes in the flat spacetime, the same scale anomalous transport can actually happen also in flat spacetime, as we will derive in this paper based on low-energy effective theory of QCD. Furthermore, we will show that it would appear dynamically in low-energy QCD.

At low-energy regime, the physics of QCD can be described by the low-lying hadron spectra including
pseudo Nambu-Goldstone bosons associated with the spontaneous breaking
of the (approximate) chiral symmetry. The chiral manifold can in principle be extremely
reduced to the one governed only by the lightest two flavors, up and down quarks,
by integrating out the heavier hadrons, so that we have only isotriplet
pions. Besides, an isosinglet scalar meson can couple to pions.
This scalar meson can be isolated and singled out to be only the one,
no matter how complicated mixing
puzzle the isosinglet scalars can have,
that may be identified as the $f_0(500)$ in the observed scalar meson
spectroscopy.

Generically, such singlet scalar couplings including self-interactions
should be parametric, not be under control.
One possible way to model singlet scalars of this kind
is to assume it to be a pseudo dilaton associated with
the spontaneous breaking of the anomalous/approximate scale symmetry,
which constrains the coupling form so as to satisfy the low-energy
theorem of the anomalous/approximate scale symmetry breaking.
In that case, the lightest isosinglet (or chiral singlet) scalar, being dilaton field,
can (at least at the classical dynamics level)
play a role of complete analogue to the scale factor
in a curved background theory.
Thus, in the QCD hadron dynamics
the scale factor $\tau(x)$ in Eq.(\ref{ano_QEDscale})
should be rephrased as the dilaton background field,
and hence, the scale anomalous transport necessarily shows up
when coupled to the EM field as well.

In contrast to the scale factor in the spacetime metric which is considered as a background configuration, nontrivial
features may come up: since
the QCD dilaton is a dynamical particle in the low-energy QCD,
the anomalous current should be dynamically generated having
the close relation with the dilaton dynamics.
Moreover, we may consider the environment of the early universe, where the hadron phase is expected to be created
in a spatially (presumably almost) homogeneous form, after the QCD phase transition and
thermalization of the hadron phase bubbles (even with feasible
inhomogeneity by a supercooling)~\cite{DeGrand:1984uq}.
The homogeneous hadron phase is thermalized with
the temperature $\sim 100$ MeV~\cite{DeGrand:1984uq,Kim:2018byy,Kim:2018knr}, at which
all hadrons already become nonrelativistic in the thermal bath with photon (but still strongly interacting each other in the kinetic equilibrium).
Then the thermal loop corrections to the dilaton dynamics can safely be neglected (with all exponentially suppressed), so
the external EM interactions would be most relevant to the dilaton,
as well as the vacuum contributions by strong interactions with pions and dilatons themselves.

The situation in heavy-ion collisions may be different where the hadron phase is expected to be inhomogeneous. However, as a first approximation, we can take the homogeneous limit as a starting point and focus on the dynamical time evolution of the system; taking the inhomogeneity of the system into account would mask the dynamical features that we will explore and thus we leave it as a future task.

We emphasize again that strong EM fields can be generated in heavy ion collision experiments~\cite{Skokov:2009qp,Deng:2012pc,Bloczynski:2013mca,Hirono:2012rt,Deng:2014uja,Voronyuk:2014rna,Huang:2015oca} and in early universe~\cite{Vachaspati:1991nm,Enqvist:1993np,Grasso:1997nx,Grasso:2000wj,Ellis:2019tjf},
where the former is due to the relativistic motion of the colliding nuclei and the smallness of the system, while the latter is, say, due to some primordial (electroweak) phase transitions of strong first order.


In this paper, we discuss a dynamic scale anomalous current induced by a dynamic dilaton
coupled to an EM field, based on a dilaton effective theory
reflecting the QCD scale anomaly in a proper way.
We find that in a homogeneous EM field background,
the dilaton potential is deformed to have a steeper and deeper well than the one without the EM filed, so that the dilaton field more promptly rolls down to the stationary point of the dilaton potential: the effective
dilaton mass (i.e. frequency of the oscillation) gets larger and
the stationary point (determining the dilaton decay constant) is shifted
to be larger.
Then, we observe that the dilaton background 
starts to oscillate in time, due to the ``kick" by the nonzero EM field, 
so that the scale anomalous transport is induced by the oscillatory dilaton field and also starts to 
oscillate.

As possible phenomenological applications,
we focus on dilepton and diphoton productions generated
from the dynamic scale anomalous transport current.
Consequently, we observe the striking peak structures of the particle productions, which are characterized by the dilaton mass.
Thus, the particle productions would be a crucial signal as an indirect detection of the dilaton in heavy ion collision experiments.

Intriguingly, the dynamic oscillatory dilaton
can create an out-of-equilibrium state due to the
nonadiabatic oscillation,
which happens in a way of the so-called tachyonic preheating mechanism~\cite{Felder:2000hj} in relation to the particle production
scenario following the inflationary epoch in the early universe.
We observe that the EM field contributes to the
tachyonic preheating as a screening effect,
so that the nonadiabaticity gets diluted by a
strong EM field.

Though we assume a dilatonic meson,
the form of the lightest isosinglet meson
coupling to EM fields
is robust,  and
necessarily couples to the scale anomaly in QCD,
as long  as the meson is isosinglet.
Hence,
the presence of the dynamic scale anomalous transport is a generic
consequence of QCD, which might trace one slice of
the thermal history of universe, and/or the created circumstance
in heavy ion collision experiments. This, thus, may open a new avenue for the anomalous transport physics in heavy ion collisions, in parallel to the dynamic particle physics that has been studied
in the inflation/preheating scenario.

Throughout this paper, we use the natural unites $\hbar=c=k_B=1$ and the most negative signature for the spacetime metric.

\section{A dilaton effective theory in EM field}
In this section, we introduce a dilaton effective theory constructed in a way parallel to the well-known chiral perturbation theory. 
In contrast to the chiral pion, however, 
we assume that the QCD dilaton cannot be exactly massless,
due to the presence of nonperturbative scale anomaly induced by the spontaneous breaking of the chiral symmetry, as in the case of nearly conformal/scale invariant gauge theories (also see footnote~\ref{nonpert-SA}).    
Note, even in that case, that 
a low-energy dilaton limit (or soft-dilaton limit) for the scale-Ward Takahashi identities
and the associated amplitudes can phenomenologically be considered, just like the Higgs-low energy theorem”~\cite{Shifman:1979eb} 
in the case of the Higgs in the standard model, which cannot have the exact Nambu-Goldstone boson limit, either. 
Though it might be crude, 
in this section, we formulate a QCD dilaton theory  
based on a spirit of the chiral perturbation theory, 
just in a sense of the Higgs low-energy theorem. 
This formulation is in sharp contrast to another dilaton effective theory approach having an exact 
scale-invariant limit and a perfect analogue of 
the chiral perturbation theory~\cite{Crewther:2013vea,Crewther:2020tgd}. 
No matter which or what approach is used,  
as long as the target scalar meson is the lightest 
and consists mainly of the lightest two-quark state, 
our main proposal will work fine,  
that is, nontrivial physics of a dynamic oscillatory scale-anomalous current is robustly present in QCD coupled with 
an EM background. 

\subsection {Lesson from the chiral perturbation theory}
To facilitate the later discussion,
for reader's convenience,
we briefly review the essential concepts of the chiral perturbation theory based on the spontaneous and explicit chiral symmetry breaking and low-energy theorem for soft pions~\cite{Gasser:1983yg,Gasser:1984gg}.

The chiral perturbation theory is constructed so as to
reproduce all Ward-Takahashi identities for
the 
chiral symmetry of the underlying QCD. 
For the chiral $SU(N_f)_L\times SU(N_f)_R$ symmetry,
among a number of the Ward-Takahashi identities, 
we have the following representative one:
\begin{eqnarray}
\partial_\mu j_5^{a\mu}=2m_f\bar q i\gamma_5 T^a q,
\label{chiral_WT}
\end{eqnarray}
where $q$ is the quark field, $m_f$ represents the quark mass (which is assumed to be the same for any flavor),
$T^a$ ($a=1,\cdots N_f$) are generators of $SU(N_f)$, and
$j_5^{a\mu}$
denotes
the axial vector current.
Once the chiral symmetry is spontaneously broken
by the chiral condensate
in the infrared vacuum ($SU(N_f)_L\times SU(N_f)_R \to SU(N_f)_{V=L+R}$),
the pions ($\pi^a$) emerge as the (pseudo) Nambu-Goldstone bosons of the chiral symmetry breaking and are coupled with the axial vector current:
\begin{eqnarray}
\langle0|j_5^{a\mu}(x) |\pi^b(p)\rangle =-if_\pi p^\mu e^{-ip\cdot x}\delta^{ab},
\label{def-fpi}
\end{eqnarray}
where $f_\pi$ is the pion decay constant.
This and Eq.~(\ref{chiral_WT}) then imply that
the nonzero quark mass explicitly breaks the chiral symmetry, so that the pions can acquire a nonzero mass.
The relation between the nonzero pion mass and the axial vector current is referred to as the partially conserved axialvector current (PCAC) relation:
\begin{eqnarray}
\langle0|\partial_\mu j_5^{a\mu}(x) |\pi^b(p)\rangle &=&-f_\pi m_\pi^2 e^{-ip\cdot x}\delta^{ab},\nonumber\\
\textrm{i.e.,} \qquad
\partial_\mu j_5^{a\mu} (x)
&=&
-f_\pi m_\pi^2 \pi^a(x)
+\cdots
.
\label{PCAC}
\end{eqnarray}
This PCAC relation can be interpreted as
the leading order relation (i.e., the so-called low-energy theorem or soft-pion theorem)
for the spontaneously broken chiral symmetry,
with respect to the perturbation in powers of $m_\pi^2$ (or quark mass in terms of
the original parameter).
The chiral perturbation theory is thus constructed
by expanding the low-energy action consistent with PCAC relation and described by
pion fields in powers of the pion mass (or quark mass), or more generically, small momenta carried by pions: the expansion parameters are thus counted as $m_\pi  \sim {\cal O}(p)$, and assumed to be
small enough compared to the chiral-symmetry breaking scale
$\sim 4 \pi f_\pi \sim 1$ GeV.

It is convenient to build the chiral Lagrangian in terms of
the chiral field $U=e^{2i\pi^a T^a/f_\pi}$.
Under the chiral transformation, $U$ is transformed as
$
U\to g_L U g_{R}^\dagger,
$
where $g_{L(R)}\in SU(N_f)_{L(R)}$.
By using $U$,
the chiral Lagrangian at leading order (i.e.,orer of ${\cal O}(p^2)$) is written as
\begin{eqnarray}
{\cal L}=\frac{f_\pi^2}{4} {\rm tr}[\partial_\mu U\partial^\mu U^\dagger]-
V(U),
\end{eqnarray}
where $V(U)$ is the potential term for pions,
which explicitly breaks the chiral symmetry and reproduces the PCAC relation in Eq.(\ref{PCAC}):
\begin{eqnarray}
V(U)= \frac{f_\pi^2 m_\pi^2}{4}{\rm tr}[
U^\dagger + 
U].
\label{pion_pot}
\end{eqnarray}
The construction of higher-order terms can be found in Refs.~\cite{Gasser:1983yg,Gasser:1984gg}.

\subsection {Scale anomaly and the low-energy theorem for the scale symmetry}


\textcolor{black}{In a way similar to the construction of chiral perturbation theory
as briefly reviewed above,
the dilaton effective theory can be established
based on Ward-Takahashi identities
for the scale symmetry.}
In the underlying QCD, taking the scale (trace) anomaly into account, the dilatation current $j_D^\mu$ satisfies
\begin{eqnarray}
\partial_\mu j^\mu_D= T^\mu_\mu =
\frac{\beta(g_s)}{2g_s}(G_{\mu\nu}^a)^2
+(1+\gamma_m) \sum_{f} m_f\bar q_f q_f,
\label{QCDAnoScale}
\end{eqnarray}
where $T_\mu^\mu$ is the trace of the energy momentum tensor,
$G_{\mu\nu}^a$ ($a=1,\cdots , 8$) is the field strength of gluons, $g_s$ is the QCD coupling constant,
$\beta(g_s)$ denotes the beta function of $g_s$,
$q_f$ represents the quark field of flavor $f$ with mass $m_f$, and
$\gamma_m$ is the anomalous dimension of the quark mass.
As noted in the beginning of the present section, 
we assume that the scale symmetry in QCD cannot be exact (even in the deeper infrared region), and, therefore, the dilaton is necessarily coupled to nonzero $T_\mu^\mu$. 
Even in that case, 
we can define the overlap amplitude between the dilaton and $T_{\mu \nu}$, 
which actually takes form similar to
the pion's in Eq.(\ref{def-fpi}):
\begin{eqnarray}
\langle 0|T_{\mu \nu}(x)|\phi(p)\rangle  
=\frac{f_\phi}{3} (p_\mu p_\nu - p^2 g_{\mu\nu}) e^{-ip\cdot x}, 
\end{eqnarray}
where $\phi$ is the Nambu-Goldstone boson called the dilaton, that has to be
an isosinglet (or chiral singlet) scalar in order to couple to the singlet current $j^\mu_D \equiv x_\nu T^{\nu \mu}$,
and $f_\phi$ is the decay constant of the dilaton.
The nonzero quark mass and the quantum gluonic corrections as shown in Eq.~(\ref{QCDAnoScale}) explicitly break the scale symmetry to 
give dilaton nonzero mass $m_\phi$: 
\begin{align}
\langle0| T_\mu^\mu (x)|\phi(p) \rangle 
= 
\langle0|\partial_\mu j^\mu_D (x)|\phi(p) \rangle
& =
-f_\phi m_\phi^2e^{-ip\cdot x}
\, \notag \\
\textrm{i.e.,} \qquad
\partial_\mu j^\mu_D (x)
& =
-f_\phi m_\phi^2\phi(x)
+\cdots
\,,
\label{pcdc_QCD}
\end{align}
with the plane-wave amplitude associated with
the one-particle dilaton state, defined as $\langle 0| \phi(x) | \phi(p) \rangle
= e^{-ip\cdot x}$.
This is called the PCDC relation (partially conserved
dilatation current),
which is analogous to the PCAC relation in Eq.~(\ref{PCAC}).
Also, consider the Ward-Takahashi identity for
the following matrix element in the low-energy limit,
\begin{align}
\lim_{q_\mu \to 0} 
\int d^4 x e^{i q\cdot x}
\langle 0| T [T_\mu^\mu (x) \,T_\nu^\nu(0)] |0 \rangle
= i
\delta_D
\langle 0| T_\mu^\mu(0) |0 \rangle
\,,  \label{WT}
\end{align}
where $\delta_D$ denotes
the infinitesimal scale/dilatation transformation
by the charge $Q_D = \int d^3 {\vec x} j_D^0(x)$, defined as
$[i Q_D, {\cal O}(x)]= \delta_D {\cal O}(x)= (d_{\cal O} + x^\nu \partial_\nu) {\cal O}(x)$, for an operator ${\cal O}$ with the scaling dimension $d_{\cal O}$.
Assuming the single-dilaton pole saturation for the left-hand side of Eq.(\ref{WT}) and applying the PCDC relation in Eq.(\ref{pcdc_QCD}) together
with the standard reduction formula,
we arrive at 
\begin{align}
\delta_D\langle 0| T_\mu^\mu (0) |0 \rangle
 = - 
 m_\phi^2 f_\phi^2
 \, \label{vac}
\end{align}
with ${\cal E}_{\rm vac.}$ being the vacuum energy.
This relation is also customarily called the PCDC relation.
Equations (\ref{pcdc_QCD}) and (\ref{vac}) are often referred to as
the low-energy theorem of the scale symmetry,
hence is generic, which all dilaton effective models based on the underlying theory QCD have to reproduce.

The presence of this kind of pseudo dilaton in QCD
has not yet been established, and is still controversial,
because of highly nontrivial infrared nonperturbativity.
As noted in the Introduction, however,
as one reference model, we shall consider the lightest isoscalar meson
(presumably corresponding to the $f_0(500)$ in the particle listing)
to be a pseudo dilaton, which may be mainly composed of
the lightest up and down quarks.
In this view, the gluon condensate term in the scale anomaly
Eq.(\ref{QCDAnoScale}) should be saturated by the up and down quark loop contributions in the nonperturbative full dynamics.

Keeping these points in our mind,
we next formulate
an effective dilaton theory constructed based on the
low-energy theorem above.  
Hereafter, for simplicity
we shall work in the chiral limit
where $m_f \to 0$.

\subsection {Dilaton effective Lagrangian }

The fundamental dynamical variable to construct the dilaton effective Lagrangian is the conformal compensator field $\chi$. Under a scale transformation $x\to x'=e^{-\sigma} x$ with the rotation angle $\sigma$, the conformal compensator field $\chi$ transforms with the scale dimension 1 as
\begin{eqnarray}
\chi(x)\to \chi'(x') =e^\sigma \chi(x)\;\;\;
{\rm or}\;\;\;
\chi(x)\to \chi'(x)=e^\sigma \chi(e^\sigma x)
.
\end{eqnarray}
The conformal compensator $\chi$ is parametrized by the dilaton field $\phi$ tagged with the decay constant of dilaton ($f_\phi$),
\begin{eqnarray}
\chi=f_\phi e ^{\phi/f_\phi}.
\end{eqnarray}
Using the conformal compensator $\chi$,
the dilaton effective Lagrangian is written as
\begin{eqnarray}
{\cal L}=
\frac{1}{2}\partial_\mu \chi \partial^\mu \chi-V(\chi),
\label{dilaton_Lag}
\end{eqnarray}
where $V(\chi)$ represents the potential term of the dilaton~\footnote{
This logarithmic form of the dilatonic scalar potential was
originally argued in~\cite{Schechter:1980ak,Salomone:1980sp,Gomm:1985ut,Migdal:1982jp,Cornwall:1984pa} based on the scale anomaly form in QCD, as
we have worked on in the present paper.
This is the minimal scale breaking form given only by the logarithmic function, and, in this sense, 
one may also refer to the one-loop radiative symmetry breaking of the Coleman-Weinberg mechanism~\cite{Coleman:1973jx}.
},
\begin{eqnarray}
V(\chi)=\frac{m_\phi^2f_\phi^2}{4}\left(\frac{\chi}{f_\phi}\right)^4\left[\log\frac{\chi}{f_\phi}-\frac{1}{4}\right],
\label{dilaton_pot}
\end{eqnarray}
in which the logarithmic potential form
corresponds to the leading order
expression in terms of the soft-scale breaking (with the explicit
scale-breaking parameter $m_\phi \ll 4 \pi f_\pi$ assumed),
by analogy to the leading-order pion potential in Eq.~(\ref{pion_pot}),
and hence,
surely reproduces the PCDC relation in Eq.(\ref{pcdc_QCD}), as will explicitly be checked below.
In the vacuum of the dilaton potential~(\ref{dilaton_pot}),
the field $\chi$ has
a nonzero vacuum expectation value,
so that the Nambu-Goldstone field $\phi$ gets null at the vacuum,
consistent with the spontaneous scale symmetry breaking
(which is also analogous to the chiral pions):
\begin{eqnarray}
\chi_0=f_\phi\,, \qquad
\phi_0 = 0 \,.
\label{st_point}
\end{eqnarray}

In terms of the ``scale-expansion" similar to the chiral
expansion for the chiral perturbation theory,
the dilaton mass $m_\phi$ is counted as ${\cal O}(p)$ (just like the pion mass $m_\pi \sim {\cal O}(p)$). 
However, as noted in the beginning of the present section, this dilaton mass 
cannot exactly be sent to zero, because of the crucial presence of 
the nonperturbative scale anomaly (see footnote~\ref{nonpert-SA}). 
Though not having the exact massless limit, a model for this dilaton can be formulated as a scale-invariant realization of the chiral perturbation theory, in which the derivative expansion starts at soft, but nonzero dilaton momentum and/or mass, along with the massless pion momentum. 
This theory is called the ``dilaton-chiral perturbation theory", which was first established in 
the literature~\cite{Matsuzaki:2013eva},
in a context of QCD with
many flavors, having 
almost scale-invariant gauge dynamics~\footnote{
Another scale-chiral perturbation 
theory has been proposed~\cite{Crewther:2013vea,Crewther:2020tgd}, 
where the nonperturbative scale anomaly 
induced by the dynamical quark mass generation 
is assumed to be washed out somehow in a renormalization-group 
invariant way.
}.

One can straightforwardly go beyond the chiral limit,
by including the current quark mass effect, like $\chi^{3-\gamma_m} m_f$, which reproduces the
quark mass term for the scale anomaly in Eq.(\ref{QCDAnoScale}).
One would then get the (next-to-leading order) chiral logarithmic
correction to the dilaton mass $m_\phi$ in Eq.
(\ref{dilaton_pot
}), in the same way as
in the dilaton-chiral perturbation theory~\cite{Matsuzaki:2013eva}.

The logarithmic potential form can also be
understood in terms of the original scale-breaking
parameter in the underlying QCD:
First note that in Eq.(\ref{dilaton_pot})
the factor $\chi^4$ and its power 4 correspond to the canonical scaling
dimension of gluon condensate in Eq.(\ref{QCDAnoScale}), which is the scale invariant part.
And then, observe that this dilaton potential $V(\chi)$ involves the explicit
breaking term, taking the log form,
which reflects
the anomalous dimension of $T_{\mu}^\mu (=d_{T_\mu^\mu} -4)$, 
arising from quantum corrections (dominated by quarks)
in the underlying QCD~\footnote{
\label{nonpert-SA}
In a sense, the dilaton field $\chi$ is interpolated by the gluon condensate induced by the quark loop.
This may roughly be understood by a naive dimensional analysis and
renormalization group property (See also Refs.~\cite{Gusynin:1987em,Hashimoto:2010nw,Matsuzaki:2015sya} for instructive
examples to analytically evaluate this quantity in the (almost) nonrunning limit of
the gauge coupling.). First, note that the
$[\beta (g_s)/g_s] (G^a_{\mu\nu})^2$ has to be finite, flavor singlet, and renormalization group
independent, and second, is now assumed to be saturated by quark loop contribution
(i.e., quark condensate). Thus, it is expected to scale like
$[\beta (g_s)/g_s] (G^a_{\mu\nu})^2 \approx \langle
[\beta (g_s)/g_s](G^a_{\mu\nu})^2\rangle \cdot (\chi/f_\phi)^{4
}$, with
$\langle
[\beta (g_s)/g_s] (G^a_{\mu\nu})^2 \rangle \sim N_c N_f m_{\rm dyn}^4/(4 \pi)^4$, in which the quark condensate gives the dynamical quark mass through
$\langle \bar{q}q \rangle_{m_{\rm dyn}} \sim N_c m_{\rm dyn}^3/(4 \pi)^2$, where $N_c=3$ and $N_f=2$.
Here the renormalization scale dependence is cancelled between
$\beta(g_s)$ and the gluon condensate operator.
Note also that $\beta(g_s)$
should correspond to
a nonperturbative beta function tied with
the infrared dimensional transmutation associated with
the nonperturbative generation of the quark condensate and the dynamical quark mass, which is essentially different from the conventional one-loop beta function in the perturbative QCD approximation.}.
The dilaton mass $m_\phi$, which becomes evident
due to the log term,
is thus related to the presence of nonzero anomalous dimension 
.

The present evaluation of the dilaton potential
may correspond to what is called the lowest-order of the scale symmetry limit~\cite{Li:2016uzn}. Other similar approaches to dilatonic
scalar potentials have recently been developed in applications to QCD or scenarios beyond the standard model.  Readers may also refer to them in, e.g, Refs.~\cite{Crewther:2013vea,Kasai:2016ifi,Hansen:2016fri,Appelquist:2017wcg,Appelquist:2017vyy,Cata:2019edh,Appelquist:2019lgk,Brown:2019ipr}.

One can easily check from the dilaton potential in Eq.~(\ref{dilaton_pot}) that
the trace anomaly in the dilaton effective model reads
\begin{eqnarray}
\partial_\mu j^\mu_D=
-\frac{f_\phi^2 m_\phi^2}{4} \left(\frac{\chi}{f_\phi}\right)^4.
\label{dilaton-mass-f}
\end{eqnarray}
Then, the matrix element for $\partial_\mu j^\mu_D$ sandwiched by the vacuum and the on-shell dilaton state with the momentum $p$ goes like
\begin{eqnarray}
\langle0|\partial_\mu j^\mu_D (x)|\phi(p) \rangle &=&
-\frac{f_\phi^2 m_\phi^2}{4} \langle0|\left(1+4\frac{\phi}{f_\phi}+\cdots\right)|\phi(p) \rangle
\nonumber\\
&=&
-f_\phi m_\phi^2e^{-ip\cdot x}.
\end{eqnarray}
This shows that the dilaton potential in Eq.~(\ref{dilaton_pot}) surely reproduces the PCDC relation in Eq.~(\ref{pcdc_QCD}) with
the chiral limit taken.
Hence, automatically, the second PCDC relation Eq.(\ref{vac}) is fulfilled
as well.


\subsection{Dilaton in EM field}
In this subsection, we shall incorporate the EM field into the dilaton effective Lagrangian in Eq.~(\ref{dilaton_Lag}).

Once the underlying QCD is coupled with an EM field,
the EM term shows up in the trace anomaly and enters the right-hand side of Eq.~(\ref{QCDAnoScale}) as (with
the chiral limit taken)
\begin{eqnarray}
\partial_\mu j^\mu_D=
\frac{\beta(g_s)}{2g_s}(G^a_{\mu\nu})^2+\frac{\beta(e)}{2e}F_{\mu\nu}^2
\label{STAwEM}
\end{eqnarray}
where
$e$ is the EM coupling constant,
$\beta(e)$ is the beta function of $e$ and
$F_{\mu\nu}$ is the field strength of the EM field with
$F^{0i}=-E^i$ and $F^{ij}=-\epsilon^{ijk}B^k$.


We suppose the dilaton effective theory to be induced
from the underlying QCD, by integrating out the
quarks and gluons, in which quarks get the dynamical mass on the order
of QCD scale ($\sim 4 \pi f_\pi$).
Since the QCD dilaton is assumed to consist mainly of
the chiral-singlet component of up and down quark bilinear,
i.e., $\bar{u}u + \bar{d}d$, the coupling to EM field
should arise from loop corrections by quarks.
Evaluation of this coupling actually involves
nonperturbative issues. The best way to qualitatively do it
is, however, to simply invoke a generic dilatonic scaling,
as if the QCD dilaton were an elementary scalar responsible for
the EM scale anomaly, so that
the dilaton field acts like a renormalization ``factor" for
the EM field, along with its beta function.
According to pertubative calculation of a dilaton-like particle
coupling to diphoton (say, for the standard-model Higgs case, that called the Higgs low-energy theorem~\cite{Shifman:1979eb}),
this evaluation could be justified when the dilaton mass is much
smaller than the dynamical quark mass scale, i.e.
the dialton can have the ``soft-dilaton" limit.
Here we will assume this limit (what we call ``dilaton low-energy theorem") to work fine,
though the expected mass hierarchy between the dilaton mass and the
dynamical (constituent) quark mass is of order one, $m_{f_0(500)}/m_{\rm dyn} \sim 500\,{\rm MeV}/300\, {\rm MeV}$.

In that case,
the effective potential of the dilaton field can be written as
\begin{eqnarray}
V(\chi)=\frac{m_\phi^2f_\phi^2}{4}\left(\frac{\chi}{f_\phi}\right)^4\left[\log\frac{\chi}{f_\phi}-\frac{1}{4}\right]
-\log\left(\frac{\chi}{f_\phi}\right)
\frac{\beta(e)}{2e}F_{\mu\nu}^2,
\label{DpotAddEM}
\end{eqnarray}
including the dilaton field coupling to the EM field
via the trace anomaly with the dilaton low-energy theorem assumed. Note also that
the added term (the second term) does not come along with
the $\chi^4$ factor, contrast to the gluonic term (the first term),
because the QCD dilaton is not interpolated by photon.
From this potential, the stationary condition for $\chi=\chi_0$
involving the EM
effects can be read as
\begin{eqnarray}
\frac{m_\phi^2}{f_\phi^2}\chi^3_0\log\frac{\chi_0}{f_\phi}
-\frac{1}{\chi_0}\frac{\beta(e)}{2e}F_{\mu\nu}^2=0.
\end{eqnarray}
By solving the stationary condition,
we find the homogeneous dilaton condensate to be
\begin{eqnarray}
\chi_0=f_\phi\left[\frac{2D}{W(2D)}\right]^{1/4},
\end{eqnarray}
where
\begin{eqnarray}
D=
\frac{1}{m_\phi^2 f_\phi^2}\frac{\beta(e)}{e}F_{\mu\nu}^2
\end{eqnarray}
with $W$ being the principal branch of the Lambert $W$ function.
The stationary point gets the EM field dependence to be shifted from the one without the EM field, as will clearly be seen below.

\subsection{Time-dependent dilaton background: ``kick" by EM field}

As noted in Introduction,
in the thermal environments created
in heavy ion collisions or early universe, strong EM fields would exist. The QCD dilaton field would be significantly coupled to the EM-field background and turns to be strongly dynamic in time. In this subsection, we discuss the time-dependent but spatially homogeneous
dilaton background $\chi_0(t)$ coupled to a homogeneous and constant EM field applied at $t=0$. We leave the study of the case of a spacetime dependent dilaton background in future. Thus, for our purpose, we add the time component of the canonical kinetic term to the dilaton effective potential, 
\begin{eqnarray}
{V}(\chi) \to
-\frac{1}{2}\partial_t \chi_0 \partial_t\chi_0
+\frac{m_\phi^2f_\phi^2}{4}\left(\frac{\chi_0}{f_\phi}\right)^4\left[\log\frac{\chi_0}{f_\phi}-\frac{1}{4}\right]
-\log\left(\frac{\chi_0}{f_\phi}\right)
\frac{\beta(e)}{2e}F_{\mu\nu}^2.
\end{eqnarray}
From this effective ``potential", \textcolor{black}{the equation of motion for} $\chi_0(t)$ reads
\begin{eqnarray}
\partial_t^2\chi_0
+
\frac{m_\phi^2}{f_\phi^2}\chi^3_0\log\frac{\chi_0}{f_\phi}
-\frac{1}{\chi_0}\frac{\beta(e)}{2e}F_{\mu\nu}^2=0.
\label{st_condition_Tcon}
\end{eqnarray}
We set the initial condition such that the dilaton sits
at the stationary point in the absence of the EM filed:
\begin{eqnarray}
\chi_0(t=0)=f_ \phi,\;\;\;
\frac{\partial \chi_0(t)}{\partial t}\Biggl{|}_{t=0}=0,
\label{initial_con}
\end{eqnarray}
and monitor a ``kick" on the dilaton by the introduced
EM field, out of the vacuum without the EM field in Eq.~(\ref{st_point}).

We consider a moderately weak EM field effect, so that
the ``kick" on the $\chi_0(t)$ is perturbative.
In that case,
we can analytically solve Eq.~(\ref{st_condition_Tcon}) with the initial condition in Eq.(\ref{initial_con}), by introducing
the linear shift $\phi_0(t)$ like 
\begin{eqnarray}
\chi_0=f_\phi e^{\phi_0/f_\phi}
=
f_\phi \left[1+\frac{\phi_0}{f_\phi} +{\cal O}\left(\frac{\phi_0}{f_\phi}\right)^2   \right].
\end{eqnarray}
Then the equation of motion for $\phi_0$ can be written as
\begin{eqnarray}
\partial_t^2\phi_0+\left( m_\phi^2+ \frac{\beta(e)}{2e}\frac{F_{\mu\nu}^2}{f_\phi^2}\right)\phi_0-
\frac{\beta(e)}{2e}\frac{F_{\mu\nu}^2}{f_\phi}=0.
\end{eqnarray}
From Eq.~(\ref{initial_con}), the initial condition for $\phi_0$ can be read as
\begin{eqnarray}
 \phi_0(t=0)=0,\;\;\;
\frac{\partial \phi_0(t)}{\partial t}\Biggl{|}_{t=0}=0.
\end{eqnarray}
We thus find the analytical solutions for the weak-field ``kick": 
\begin{eqnarray}
\phi_0(t)&\approx&f_\phi D
\sin^2 \frac{m_\phi^{\rm eff}(D) t}{2}
%
%
,\nonumber\\
\chi_0(t)&\approx&f_\phi
\left(
1+D
\sin^2 \frac{m_\phi^{\rm eff}(D) t}{2}
\right)\,,
\label{chi_approx}
\end{eqnarray}
with 
\begin{eqnarray}
m_\phi^{\rm eff}(D) &=& \sqrt{1+\frac{D}{2}} m_\phi \,, \notag \\
D
&=& \frac{\beta(e)}{e}\frac{F_{\mu\nu}^2}{m_\phi^2 f_\phi^2} \ll1.
\label{eff-mass}
\end{eqnarray}
It is clear to see that even a weak EM field can drive the dilaton background field $\chi_0(t)$ in Eq.~(\ref{chi_approx}) to oscillate. The external EM field supplies an energy to the dilaton potential, and therefore ``kicks" the dilaton background to oscillate, which is used to stay at
the vacuum. We give more detailed discussions in the following sections~\footnote{Another interesting point to notice is that
the dilaton mass dictated by the oscillation frequecy
is now ``screened" by the EM field environment (Eq.(\ref{eff-mass})).
This originates from the trace anomaly induced from
the external EM field,
which the dilaton coupled to, and hence,
is a view similar to the chameleon mechanism~\cite{Khoury:2003aq,Khoury:2003rn} that have extensively been
studied in the field of cosmology.}.

\section{Dynamic scale anomalous transport in EM field}
As long as the dynamics is disregarded,
the QCD dilaton that we presently focus on is thought to be
a generic dilatonic scalar, as it should be followed by our assumption
made in the previous section,
hence can be viewed as
just a scale factor in a classical Maxwell theory with the metric curved by
the dilaton. In fact,
one can easily check that
the dilaton field can completely be transformed away by a Weyl transformation:
$\eta_{\mu\nu} \to e^{\phi(x)/f_\phi} \eta_{\mu\nu}$, and a field redefinition
of the EM field, $A_\mu(x) \to e^{ \phi(x)/f_\phi} A_\mu(x)$.
This transformation should be anomalous when quarks coupled to the EM field
are introduced,
leaving a scale (Weyl) anomaly scaled by the beta function arising from
the quark loop, that exactly corresponds to
the EM scale anomaly term in Eq.(\ref{STAwEM}) (See, e.g., Refs.~\cite{Fujikawa:1980vr,Fujikawa:1980rc,Fujikawa:1993xv,Katsuragawa:2016yir,Kamada:2019pmx} for
Weyl anomaly and discussion on the frame equivalence applied to particle physics).
Thus, including this EM scale anomaly,
the dilaton effective theory without the dynamics
is Weyl equivalent to the Maxwell theory in a curved spacetime.
Therefore, it is obvious that
we can reproduce the scale anomalous transport current derived
in Refs.~\cite{Chernodub:2016lbo,Chernodub:2017bbd,Chernodub:2017jcp,Zheng:2019xeu} for the Maxwell theory
in the curved background, as briefly noted in the Introduction.
This may be called an ``{\it equivalence theorem}" for the scale anomalous transport physics, which completely fixes the transport current form.

However, the dynamics of the scale factor (i.e., presently the QCD dilaton)
would provide a discriminating transport physics.
In this section, we shall discuss a couple of
phenomenological implications
arising from a dynamic scale anomalous
transport current induced by the dynamic oscillatory dilaton,
based on the oscillation profile in Eq.(\ref{chi_approx})
in a moderately weak EM-field background.
Those dynamical features would be characteristic to
the spatially homogeneous hadron phase, as elaborated
in the previous section, and could be potentially probed
in heavy ion collisions and/or some thermal history in early universe.

We begin by deriving the universal scale anomalous current form
followed by the equivalence theorem.
The derivation is actually given in a way similar to the anomalous
transports due to the axial anomaly~\footnote{
The QCD $\theta$-potential term,
\begin{eqnarray*}
V_\theta=-\int d^4x \theta(x)\frac{g_s^2}{32\pi^2}
\epsilon^{\mu\nu\rho\sigma}{\rm tr}[G_{\mu\nu}G_{\rho\sigma}]
\end{eqnarray*}
 should be related to the axial anomaly for the isosinglet axial current $j_A^\mu$ in the presence of an
 EM field:
\begin{eqnarray*}
\partial_\mu j_A^\mu=\sum_f2m_f\bar q_f i\gamma_5q_f
-\frac{g_s^2 N_f}{16\pi^2}
\epsilon^{\mu\nu\rho\sigma}{\rm tr}[G_{\mu\nu}G_{\rho\sigma}]
-\frac{N_c e^2}{32\pi^2}\epsilon^{\mu\nu\rho\sigma}F_{\mu\nu}F_{\rho\sigma}{\rm tr}[Q^2],
\end{eqnarray*}
where $Q$ is the electric charge matrix. By inserting this anomaly form into $V_\theta$,
the $\theta$ term goes like
\begin{eqnarray*}
V_\theta
&=&\int d^4x \theta(x)\frac{1}{2N_f}\partial_\mu j_A^\mu
-\int d^4x \theta(x)\frac{1}{N_f}\sum_fm_f\bar q_fi\gamma_5 q_f\nonumber\\
&&
+\int d^4x\theta(x)\frac{N_ce^2}{64\pi^2 N_f}
\epsilon^{\mu\nu\rho\sigma}F_{\mu\nu}F_{\rho\sigma}{\rm tr}[Q^2].
\end{eqnarray*}
Varying $V_\theta$ with respect to $A_\mu$, the last term gives a chiral anomalous EM current,
\begin{eqnarray*}
j^\mu=\frac{e^2 N_c }{16\pi^2 N_f}\epsilon^{\mu\nu\rho\sigma}
\partial_\nu \theta F_{\rho\sigma}
{\rm tr}[Q^2].
\end{eqnarray*}
For a time-dependent but homogeneous $\theta$, say, $\theta=N_f\mu_A t$ with $\mu_A$ being a constant (called chiral chemical potential), this current reproduces the well-known chiral magnetic effect~\cite{Kharzeev:2007jp,Fukushima:2008xe}. If one promotes $\theta$ to being a dynamic field (the axion) and adds the corresponding kinetic and potential terms into $V_\theta$, one would thus be able to study the dynamic chiral anomalous transport in EM field in a manner similar to the dynamic scale anomalous transport discussed in the main text.
}.
From the dilaton potential in Eq.~(\ref{DpotAddEM}),
we readily find the anomalous EM current, to get
\begin{eqnarray}
 j^\mu(x)
=\frac{\delta }{\delta A_\mu(x)}\int d^4y { V}(\chi)
\,.
\end{eqnarray}
As a function of the oscillatory dilaton background $\chi_0$
in the homogeneous space,
this current is evaluated as
\begin{eqnarray}
\langle j^\mu(x)\rangle&=& -F^{\mu\nu}
\left[\partial_\nu\log\left(\frac{\chi_0}{f_\phi}\right)
\right]\frac{2\beta(e)}{e}
+
\left(\partial_\nu F^{\nu\mu}\right)\frac{2\beta(e)}{e}\log\left(\frac{\chi_0}{f_\phi}\right).
\label{anocurrent}
\end{eqnarray}
This current form is equivalent to the one derived from Maxwell theory in
a curved spacetime with the scale factor $\tau=\phi/f_\phi$ (see Eq.(\ref{ano_QEDscale})), as it should be.

Since the present target system is assumed spatially homogeneous, we may consider a mean field approximation for the EM field and assume a constant EM field,
$ F_{\mu\nu}(x) \equiv F_{\mu\nu}$.
In this case,
the surviving component of the anomalous current goes like
\begin{eqnarray}
\langle j^i(t)\rangle=
-E^i
\left[\frac{d}{dt}\log\left(\frac{\chi_0(t)}{f_\phi}\right)
\right]\frac{2\beta(e)}{e}.
\label{ano_current}
\end{eqnarray}
Using Eq.(\ref{chi_approx}), in a weak EM field,  we have 
\begin{align}
\langle j^i(t)\rangle\approx
-E^i
\cdot m_{\phi}^{\rm eff}(D) 
\frac{\beta(e)}{e}
D
\sin (m_\phi^{\rm eff}(D) t)
.
\label{ano_current-approx}
\end{align}
Thus, this dynamic current is created in the vacuum governed by
the oscillatory dilaton configuration~\footnote{We note that despite the current (\ref{ano_current-approx}) has the form of the usual Ohm current in a conducting medium, its origin is completely different from the Ohm's law, as it arises from the scale anomaly in a dilaton-condensed vaccum rather than in a thermal or dense medium and the current (\ref{ano_current-approx}) can either be along or opposite to the direction of the applied $\vec E$ field meaning that the applied EM field could either supply or extract energies from the time-dependent 
dilaton background.}.



\subsection{Dilepton production from the dynamic oscillatory vacuum}

In this subsection, we discuss the dilepton (specifically, $e^+e^-$) production, generated
from the vacuum having the dynamic scale anomalous current in Eq.(\ref{ano_current}),
induced by the oscillatory dilaton background~\footnote{A related calculation for the dilepton production due to scale anomaly in quark-gluon plasma in the hydrodynamic regime was studied in Ref.~\cite{Basar:2012bp}.}.
To this end, we first introduce a dynamical photon field in $A_\mu$ as
\begin{eqnarray}
A_\mu=\bar A_\mu+\tilde A_\mu,
\end{eqnarray}
 where $\bar A_\mu$ is the background field of the EM field and $\tilde A_\mu$ is the dynamical photon field.
Looking at the dilaton effective potential in Eq.~(\ref{DpotAddEM})
with the anomalous current in Eq.~(\ref{anocurrent}) or Eq.(\ref{ano_current-approx}),
one can find the effective interaction between the dynamical photon and the anomalous current,
\begin{eqnarray}
{\cal L}_{\gamma j}=-\tilde A_\mu \langle j^\mu\rangle.
\end{eqnarray}
From this effective interaction,
we see that
the anomalous current induces the dilepton production through the
minimal EM interaction term between electron ($\psi$) and photon fields,
${\cal L}_{\rm int}=e\bar\psi A_\mu \gamma^\mu \psi $: 
\begin{eqnarray}
\langle e^-(p,s) e^+(q,s') |\Omega\rangle
&=&
\langle e^-(p,s) e^+(q,s')|
 \left[i\int d^4x e\bar\psi \tilde A_\mu \gamma^\mu \psi\right]
\left[-i\int d^4 y\tilde A_\nu \langle j^\nu\rangle\right]
|0\rangle\nonumber\\
&=&e\left[ \bar u^s(p)\gamma^\mu v^{s'}(q) D_{\mu\nu}^{(\gamma)}(p+q)  \right]
\int d^4x \langle j^\nu(t)\rangle e^{i(p+q)\cdot x},
\end{eqnarray}
where $D_{\mu\nu}^{(\gamma)}$ is the photon propagator, $s(s')$ is the electron (positron) spin and
$|\Omega\rangle$ represents the vacuum corresponding to the stationary point of the  dilaton potential at $\chi_0(t)$, which has created the anomalous current $j_\mu(t)$.
By using Eq.(\ref{ano_current-approx}),
this production amplitude is further evaluated to be
\begin{eqnarray}
\langle e^-(p,s) e^+(q,s') |\Omega\rangle
&=&e\left[ \bar u^s(p)\gamma^\mu v^{s'}(q) D_{\mu\nu}^{(\gamma)}(p+q)  \right]
\left(\bar F^{\nu 0}
\frac{1}{i}\frac{\beta(e)}{e} \frac{D}{2}  m^{\rm eff}_\phi(D)
\right)\nonumber\\
&&\times (2\pi)^4\delta\left(
m_\phi^{\rm eff}(D) -p_0-q_0\right) \delta^{(3)}(\vec p+\vec q).
\label{dilepton-amplitude}
\end{eqnarray}
This nonzero production amplitude
shows that the back-to-back dilepton pair is emitted from the anomalous current.
Remarkably enough, the dilepton-production distribution has an intrinsic peak
at $m_{e^+e^-}= 
m_\phi^{\rm eff}(D)$, when the two electron energies are reconstructed
to be the dilepton invariant mass $m_{e^+ e^-}$
as $p_0=q_0=m_{e^+ e^-}/2$.
In the ordinary QCD hadron physics,
this kind of peak structure eminent by a dilaton
mass (i.e. isosinglet scalar meson mass)
cannot be realized
in the dilepton production,
because of the border resonant property, 
and also the coupling to leptons is so extremely small 
that the peak size should be small as well~\footnote{
The isosinglet $f_0$ meson couplings to dilepton would 
arise from the electroweak interactions at two-loop level, 
the size of which would then be highly suppressed 
by some power of the QCD hadron scale over the electroweak scale, 
compared to the corresponding coupling strength produced from 
the dilaton-oscillatory vacuum in Eq.(\ref{dilepton-amplitude}) 
with a typical EM field strength of order of $m_\pi$ taken into account.  
}. 
Therefore, this anomalous dilepton production might be a distinct phenomenological consequence which provide a novel means to probe the presence of the dynamic scale anomalous transport and in turn the presence of QCD dilaton in heavy-ion collision experiments.


Computing the square of Eq.(\ref{dilepton-amplitude}), we further get
the dilepton production rate in unit of the phase space in the $e^+ e^-$
plane,
\begin{eqnarray}
&&\sum_{s,s'}|\langle e^-(p,s) e^+(q,s') |\Omega\rangle|^2
\nonumber\\
&&=4\left[e^4\left\{
p^{\mu} q^\nu+p^{\nu} q^\mu-g^{\mu\nu}(p\cdot q+m_e^2)
\right\}D_{\mu\rho}^{ (\gamma)}(p+q) \bar F^{\rho0}
\left(D_{\nu\sigma}^{(\gamma)}(p+q)
\right)^\dagger \bar F^{\sigma0}
\right]\nonumber\\
&&
\times \left(
\frac{\beta}{e}\frac{D}{
2
}
m_\phi^{\rm eff}(D)
\right)^2
\left((2\pi)^4\delta\left( 
m_\phi^{\rm eff}(D) -p_0-q_0\right) \delta^{(3)}(\vec p+\vec q) \right)^2,
\end{eqnarray}
and, integrating over the phase space,
arrive at the production rate of the dilepton pair per unit volume, $\Gamma_{e^+ e^-}$,
\begin{eqnarray}
\Gamma_{e^+e^-}\equiv\frac{N_{e^+ e^-}}{V_4}&=&
\vec E^2 \cdot
\frac{e^2\beta^2}{8\pi}
D^2
\left(
1-\frac{1}{3}\frac{[m_\phi^{\rm eff}(D)]^2-4m_e^2}{[m_\phi^{\rm eff}(D)]^2}
\right)\frac{\sqrt{[m_\phi^{\rm eff}(D)]^2-4m_e^2}}{m_\phi^{\rm eff}(D)}
\notag \\
&\simeq&
\frac{\beta^2}{4 \pi}(e\vec E)^2 \cdot \frac{D^2}{3}
\,,
\label{pro_diplepton}
\end{eqnarray}
where
$N_{e^+e^-}$ is the number of dilepton pair, $V_4$ is the spacetime volume,  $V_4=(2\pi)^4 \delta^{(4)}(p)|_{p=0}$,
and in the second line only the leading term has been kept
with $m_\phi \gg m_e$ taken into account.

\textcolor{black}{It is worth comparing the above dilepton production induced by scale anomaly with the Schwinger pair production in a pure electric field~\cite{Schwinger:1951nm}. The Schwinger pair production rate is given by $\Gamma_{\rm Sch}=[(e\vec E)^2/4\pi^3]\exp(-\pi m_e^2/|e\vec E|)$. First, the scale anomalous pair production is a perturbative effect for arbitrarily strong electric field (our calculation is, however, performed for $eF_{\mu\nu}$ weaker than $m_\phi f_\phi$) while the Schwinger pair production is a nonperturbative tunneling effect which can occur only when the applied electric field is larger than $m_e^2$ as can be easily seen in $\Gamma_{\rm Sch}$ which is exponentially suppressed when $|e\vec E|\ll m_e^2$. Second, as we have noted, the dilepton invariant mass would peak at $m_\phi^{\rm eff}(D)$ in the scale anomalous production while there is no such feature in the Schwinger pair production; instead, the Schwinger pair with invariant mass larger than $|e\vec E|$ would be exponentially suppressed. Thus, though for electric field in the regime $m_\phi^{\rm eff} f_\phi\gg |e\vec E|\gg m_e^2$ the total dilepton production would be dominated by Schwinger mechanism, the special invariant mass distribution of the pair would make the scale anomalous dilepton production distinguishable.}

\subsection{Diphoton production from the dynamic oscillatory vacuum}

Similarly, the EM interaction of the dilaton field in Eq.~(\ref{STAwEM}) induces the diphoton production from the oscillatory dilaton background~\footnote{
Single photon production is kinematically 
disallowed, because of the momentum conservation. 
}.
In the weak EM fields, by using Eq.~(\ref{ano_current-approx}),
the diphoton production amplitude goes like 
(in the first-order approximation of the perturbation in $\beta(e)$) 
\begin{eqnarray}
\langle \epsilon^{(i)} ({\vec p}) \epsilon^{(j)} ({\vec q})  |\Omega\rangle
&=& i\frac{\beta(e)}{e}
\frac{D}{2}
\left\{
- (p\cdot q) (\epsilon^{(i)} ({\vec p}) \cdot \epsilon_{(j)} ({\vec q}))
+(q \cdot  \epsilon^{(i)} ({\vec p})) (p \cdot \epsilon^{(j)} ({\vec q}))
\right\}\nonumber\\
&&\times
(2\pi)^4\delta\left(
m_\phi^{\rm eff}(D) -p_0-q_0\right) \delta^{(3)}(\vec p+\vec q),
\label{diphoton-amp}
\end{eqnarray}
where $\epsilon^{(i)} ({\vec p})$ is the polarization of the photon having
the momentum ${\vec p}$, and
we have used the on-shell-photon condition, $p^2=q^2=0$ and $q^\mu \epsilon^{(i)}_\mu ({\vec q})=p^\mu \epsilon^{(i)}_\mu ({\vec p})=0$,
and $p_0>0,\; q_0>0$.
It is noteworthy
that the diphoton pair production induced by the oscillating dilaton background
is also observed as a back-to-back photon emission, which will yield
the sharp peak structure in the diphoton invariant mass distribution at $m_{\gamma\gamma} = m_\phi^{\rm eff}(D)$.
Compared to the kinematics in the ordinary diphoton mass distribution
to which some isosinglet scalar mesons can contribute through
the same EM-scalar coupling of $\phi-\gamma-\gamma$ form,
as in Eq.(\ref{STAwEM}),
the ordinary diphoton production amplitude
involves the scalar meson exchange, so
should include the resonance structure, which
is obviously quite broad (compared with delta function)
for all the isosinglet scalar candidates.
In contrast, the production amplitude in Eq.(\ref{diphoton-amp})
is sharply peaked at $m_{\gamma\gamma} = m_\phi^{\rm eff}(D)$,
so is clearly distinguishable from the ordinary scalar-diphoton production
process, by the intrinsic kinematics feature. The ultra-peripheral heavy-ion collisions may provide an environment in which this diphoton emission from dynamic oscillatory anomalous current could possiblly be measured.

Taking the square of the amplitude,
\begin{eqnarray}
&& \sum_{i,j}|\langle \epsilon^{(i)} ({\vec p}) \epsilon^{(j)} ({\vec q})
|\Omega\rangle|^2
\nonumber\\
&&=
\frac{1}{8}\left(\frac{\beta (e)}{e}\frac{\sqrt{F_{\mu\nu}^2} }{f_\phi}\right)^4 \left((2\pi)^4\delta\left( 
m_\phi^{\rm eff}(D) -p_0-q_0\right) \delta^{(3)}(\vec p+\vec q) \right)^2,
\end{eqnarray}
and integrating over the phase space, the diphoton production rate per unit volume, $\Gamma_{\gamma\gamma}=N_{\gamma\gamma}/V_4$, is evaluated as
\begin{eqnarray}
\Gamma_{\gamma\gamma}
\simeq
\frac{1}{ 64 \pi} [m_\phi^{\rm eff}(D)]^4  \frac{\beta^2}{e^2}
D^2
.
\label{pro_diphoton}
\end{eqnarray}
Taking the ratio of this to $\Gamma_{e^+e^-}$ in Eq.(\ref{pro_diplepton}),
we find, to the leading order of $D$, 
\begin{align}
 \frac{\Gamma_{\gamma\gamma}}{\Gamma_{e^+e^-}}
 \simeq
 \frac{3 b_1}{ 128 \pi^2 e^2} \frac{m_\phi^2}{f_\phi^2}\cdot \frac{1}{D}
 \sim  
 \frac{1}{e^2}
 \left(\frac{m_\phi}{500\,{\rm MeV}} \right)^2
 \left(\frac{100\,{\rm MeV}}{f_\phi} \right)^2\cdot
 \left( \frac{0.1}{D} \right)
 \,,
\end{align}
where we set $\vec{E}^2 = F_{\mu\nu}^2/2$ in Eq.(\ref{pro_diplepton}), used
Eq.(\ref{eff-mass}), and took $\beta(e)=\frac{e^3}{(4\pi)^2} b_1$ with
the perturbative one-loop beta-function coefficient arising from up and down quark loops,
$b_1 = \frac{2 N_c}{3} ((-1/3)^2 + (2/3)^2)$ with $N_c=3$.
Thus, the diphoton is more efficiently produced than the dilepton
in a weak EM field limit ($D \ll 1$).


\section{Dynamics of oscillatory scale anomalous current and dilaton}

We have so far observed that the dynamic oscillatory scale anomalous transport
current, induced from the oscillating dilaton ``vacuum",
provides intriguing anomalous productions of dilepton and diphoton,
and those production events can have the characteristic kinematics explosively
enhanced by the effective dilaton mass threshold, which
could be testable in the experiments for heavy ion collisions.
In this section, more on the intrinsic oscillatory dynamics is
explored, by focusing on the time-evolution dynamics of the dilaton, and, beyond the weak EM field approximation,
we work on numerical calculations.

We fix the dilaton mass (without the EM field background)
to the mass of the lightest isosinglet scalar meson,
$f_0(500)$,~\footnote{
The input value of $f_\phi$ would be reasonable
when the present dilaton theory is induced from
the two-flavor linear sigma model with heavier scalar mesons integrated out.
In that case, the surviving chiral-singlet scalar (arising as the radial component of the conventional
sigma and pion fields) should show up with
the wavefunction renormalization factor $f_\pi$,
which is identified as $f_\phi$ in the dilaton
effective theory. In this sense, $f_\phi$ may conservatively be
thought to be on the same order as $f_\pi$.}$^,$\footnote{
The present analysis is done in the chiral limit,
so one might think the chiral-limit $f_0(500)$ mass
can be quite different from the observed one.
We have checked the current mass effect on the $f_0(500)$
identified as the dilaton, by using the established
dilaton-chiral perturbation theory proposed in Ref.~\cite{Matsuzaki:2013eva}.
Setting the dilaton mass at the physical
pion mass ($m_\pi=$140 MeV, in the isospin symmetric limit) to be $m_\phi(m_\pi=140\,{\rm MeV})= 500$ MeV
and the physical dilaton decay constant
to be $f_\phi(m_\pi=140\,{\rm MeV}) = 100$ MeV (as in Eq.(\ref{mass-fit})),
we extract the chiral-limit dilaton mass
$m_\phi(m_\pi=0)$,
based on the chiral-extrapolation formula for
the dilaton mass including the chiral logarithmic
corrections, given in the literature,
to find $m_\phi(m_\pi=0) \simeq 478$ MeV (i.e. only about 4\%
correction from the current quark masses).
So, our input value 500 MeV in Eq.(\ref{mass-fit}) is good enough even for
the chiral-limit analysis.
Note also that the discrepancy for the dilaton decay constants, $f_\phi(m_\pi=140\,{\rm MeV})$ and $f_\phi(m_\pi=0)$, turns out to be of the same size.
}$^,$\footnote{
The selected set values for $(m_\phi, f_\phi)$
is slightly small (by about 35\%) compared to an expected size from
the (second) PCDC relation Eq.(\ref{vac}),
where the right-hand side gives $(m_\phi f_\phi)^2 \simeq (224\, {\rm MeV})^4$, while the left hand side can be evaluated by referring to the result from
the QCD sum rule~\cite{Shifman:1978bx,Shifman:1978by}:
$\langle0| T_\mu^\mu |0 \rangle
= b/8 \langle 0| \frac{\alpha_s}{\pi} [G_{\mu\nu}^a]^2) |0\rangle \simeq (347 \, {\rm MeV})^4$
($\alpha_s \equiv g_s^2/4\pi$) with
the one-loop QCD beta function coefficient (only for up and down quark loops) $b=29/3$ assumed.
However, this discrepancy may be reasonable, because
it also involves a systematic
uncertainty for the QCD sum rule, plus the chiral extrapolation
from the real-life QCD to the chiral-limit on the observed hadron cross sections
to derive the value of the gluon condensate. }
\begin{eqnarray}
m_\phi=500\,{\rm MeV},\;\;\;
f_\phi=100\,{\rm MeV}.
\label{mass-fit}
\end{eqnarray}
For the beta function $\beta(e)$, we take the perturbative one-loop result, as was done in the previous section.
The unit of the EM field is the pion mass, $m_\pi=140$ MeV.
Then, the weak EM field condition Eq.(\ref{eff-mass}) can quantitatively be understood as follows:
$D = 5/(18 \pi^2) \cdot (eF_{\mu\nu})^2/(m_\phi f_\phi)^2 \ll 1$,
namely, $(eF_{\mu\nu})^2 \ll (4.0\, m_{\pi})^4$, for
$m_\phi =500$ MeV and $f_\phi = 100$ MeV, with the one-loop beta function
coefficient ($b_1$) as above.


\subsection{Dynamic oscillatory scale anomalous current
in EM field}

\begin{figure}[t]
  \begin{center}
   \includegraphics[width=12cm]{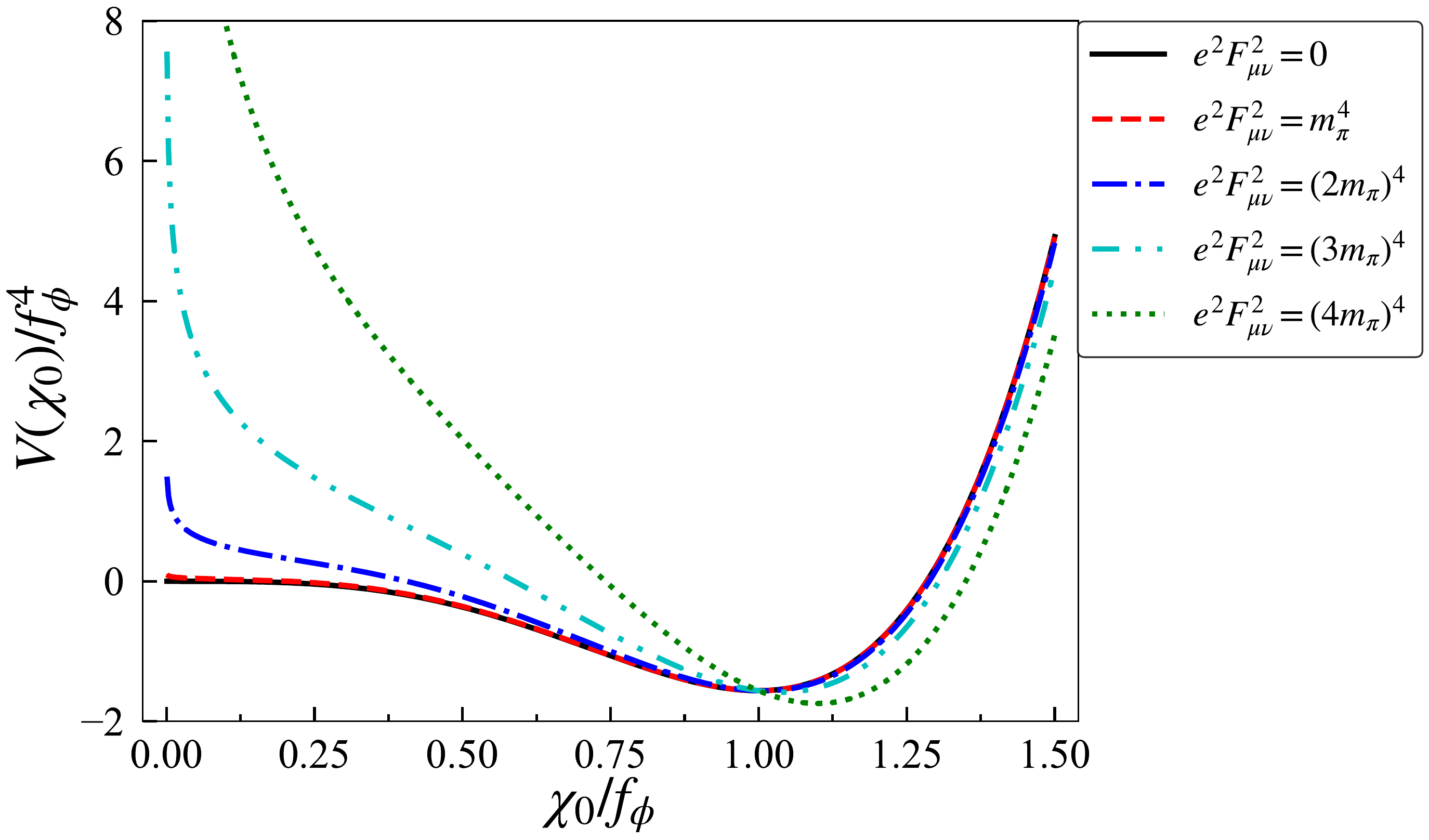}
  \end{center}
 \caption{
Plot of the dilaton potential
in the EM field, as a function of $\chi_0/f_\phi$,
which sketches the screening effect by the created EM-field barrier.
The strength of the EM field has been varied in unit of $m_\pi^2$. }
 \label{pot_cal}
\end{figure}

First of all, let us examine
the EM effect on the dilaton potential.
See Fig.~\ref{pot_cal}.
As the EM field strength increases,
the dilaton potential
gets stabilized to be pulled down into a steeper and deeper potential well than the previous vacuum.
This deformation is manifestly due to the screening effect by the EM field:
the effective dilaton mass, observed as the curvature of the potential minimum,
is indeed enlarged (see also Eq.(\ref{eff-mass}) in the weak field limit),
and the saddle point at the origin is kicked and lifted up because of
creation of the EM screening barrier, which makes
the dilaton trapped at the potential minimum.
This picture helps us to easily understand that,
by the EM field,
the dilaton rolls down to the stationary point of the dilaton potential, which used to stay at the vacuum without the EM field, so starts to oscillate.
The Fig.~\ref{pot_cal} will also give an intuitive interpretation on the dynamics of
the oscillatory dialton background, described below.
Note that in the weak EM field where $e^2 F_{\mu\nu}^2 \ll (4.0\, m_{\pi})^4$,
the screening effect is still small enough to keep the
original dilaton potential shape, so the dilaton cannot be
trapped at the potential minimum.


\begin{figure}[t]
  \begin{center}
   \includegraphics[width=12cm]{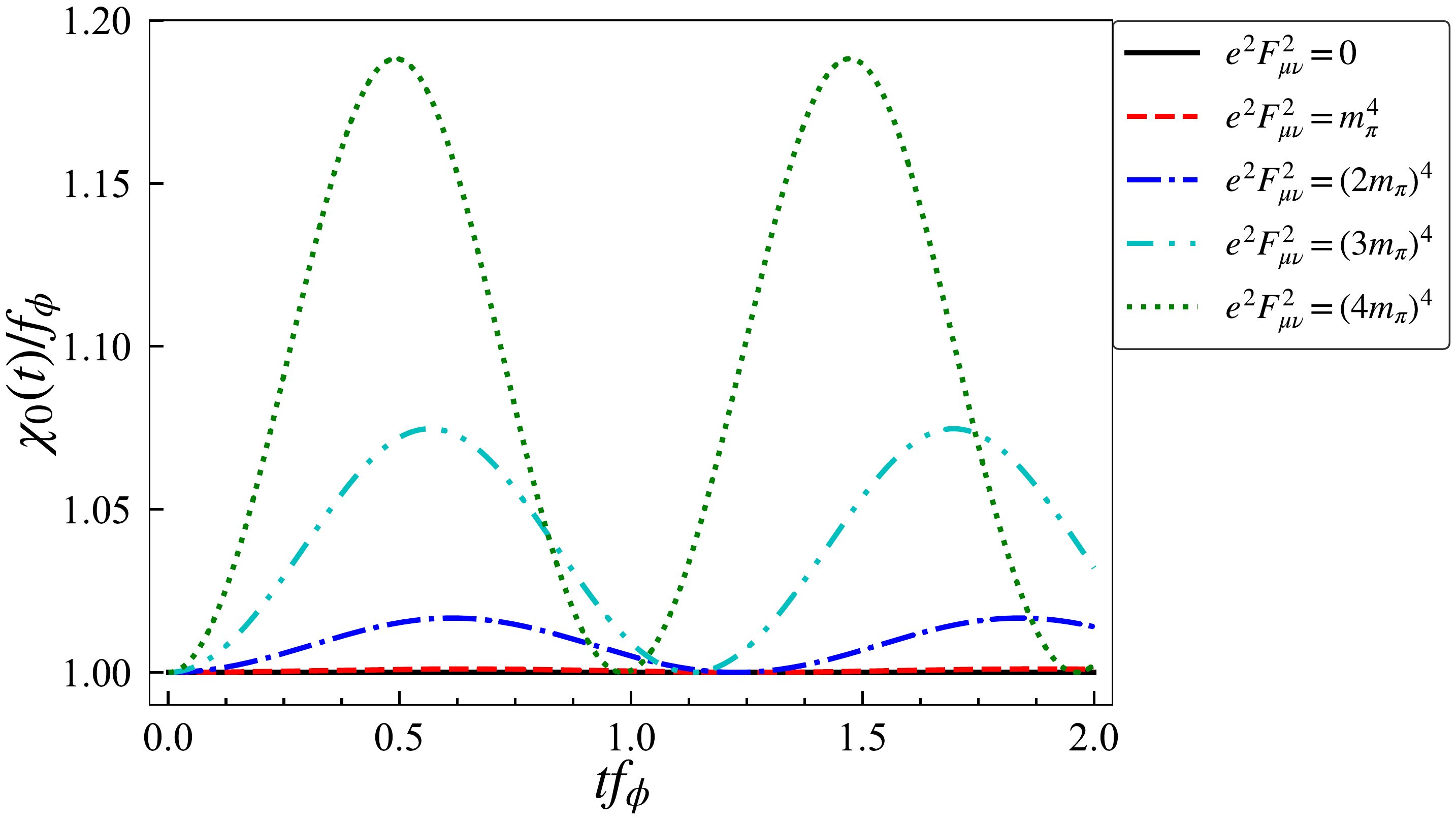}
  \end{center}
 \caption{
Development of the oscillatory dilaton background $\chi_0(t)$ as a function of the normalized
time scale $tf_\phi$
by increasing the strength of the EM field.}
 \label{Tcon_cal}
\end{figure}

Next, by numerically solving the stationary condition in Eq.~(\ref{st_condition_Tcon}) with the initial condition in Eq.~(\ref{initial_con}), the time-dependent dilaton background is evaluated.
Fig.~\ref{Tcon_cal} shows the EM-field dependence of the time-dependent dilaton background $\chi_0(t)$.
Note from the chosen initial condition in Eq.(\ref{initial_con}) that
in the absence of the EM field,
the $\chi_0(t)$ keeps constant in time, and stays at the vacuum.
By switching on the EM field, the vacuum turns to feel it, and
the stationary point is shifted to be a deeper potential minimum specifying a new vacuum,
as illustrated in Fig.~\ref{pot_cal}.
Thus the dilaton is kicked by the EM field and starts to
move to the new vacuum, and then starts to oscillate around the new vacuum.
As the EM field increases, the amplitude of the dilaton oscillation
becomes larger and larger, and the frequency of the oscillation becomes
more rapid as well.

\begin{figure}[t]
  \begin{center}
   \includegraphics[width=12cm]{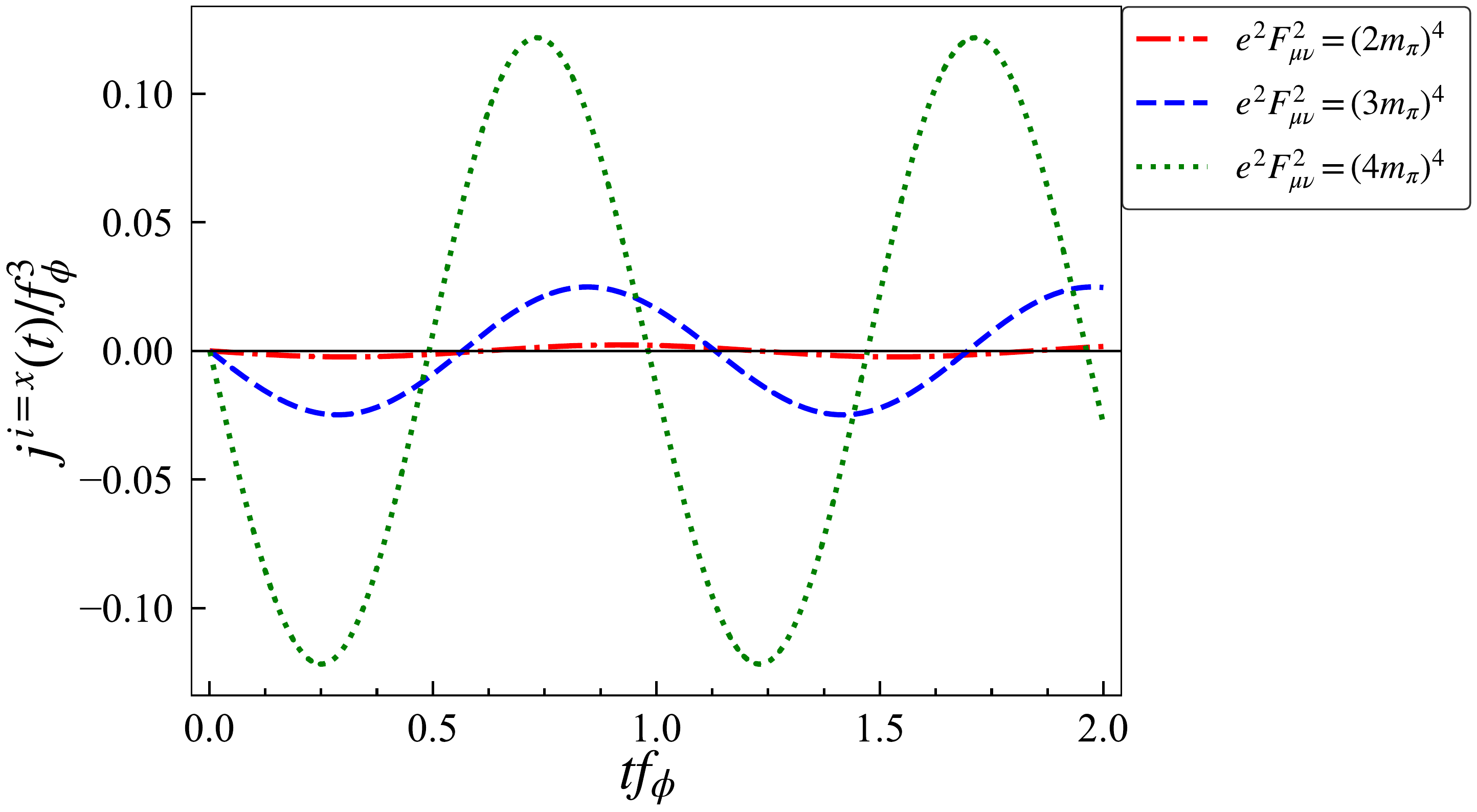}
  \end{center}
 \caption{
Time evolution of the oscillating scale anomalous current in the EM field,
plotted as a function of $tf_\phi$. }
 \label{current_cal}
\end{figure}

Now, using the time-dependent dilaton background as depicted in Fig.~\ref{Tcon_cal}, we numerically evaluate the EM dependence of the dynamic oscillatory scale anomalous current in Eq.~(\ref{ano_current}).
To numerically estimate the anomalous current, we take $E^x$ to be  $E^x=\sqrt{F_{\mu\nu}^2}/2$.
Fig.~\ref{current_cal} shows the EM dependence of the anomalous current.
It is now visible that the anomalous current oscillates because of the
oscillatory dilaton background.
As the EM fields increase, the anomalous current oscillates quicker and the oscillating amplitude becomes larger, in the same manner as the dilaton background does in Fig.~\ref{Tcon_cal}.

\subsection{Time-dependent dilaton ``mass" on the oscillatory background in EM field: nonadiabaticity versus EM screening}

\begin{figure}[t]
\begin{tabular}{cc}
 \begin{minipage}{0.5\hsize}
  \begin{center}
   \includegraphics[width=7.5cm]{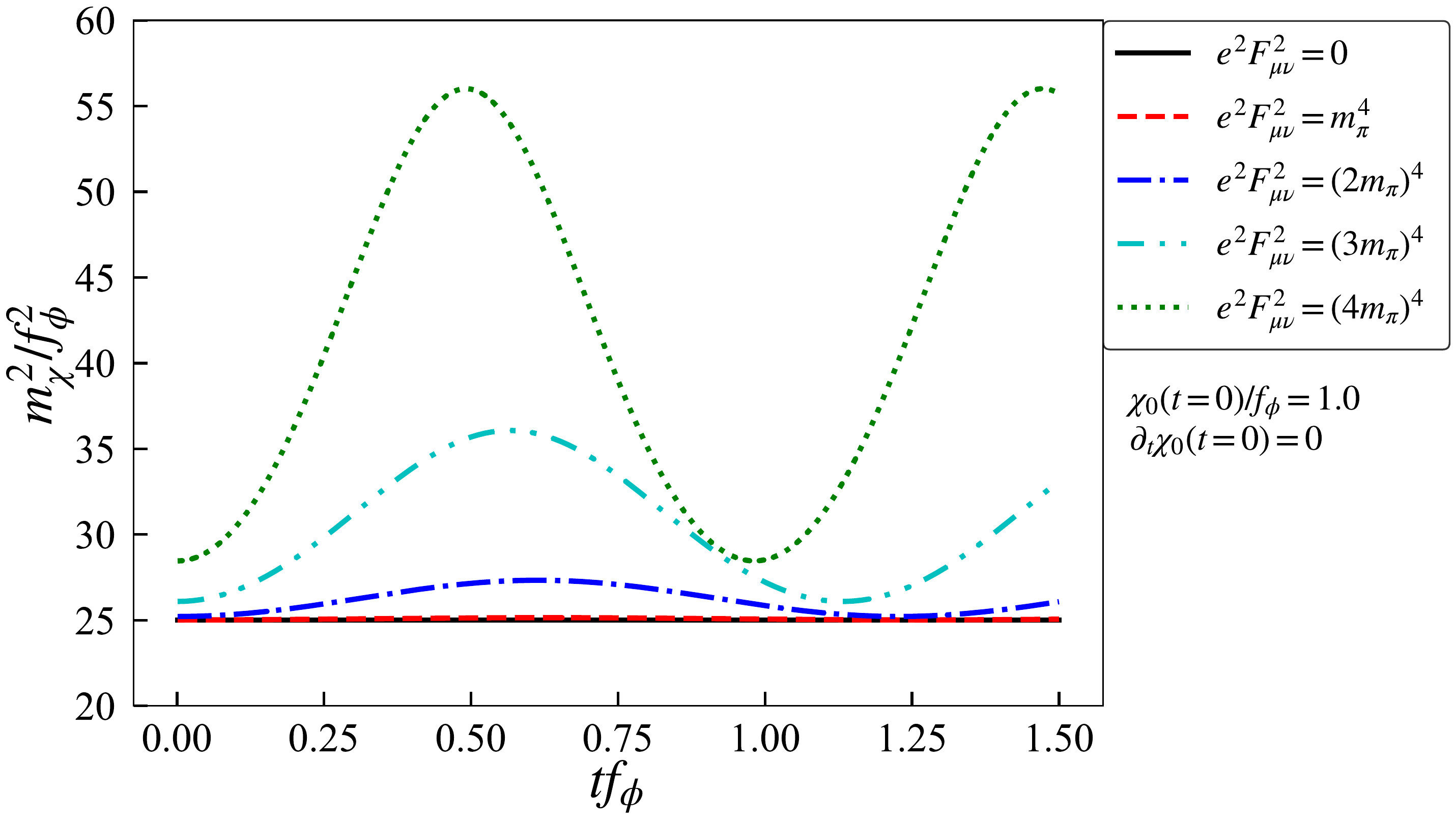}
    \subfigure{(a)}
  \end{center}
 \end{minipage}%
 \begin{minipage}{0.5\hsize}
  \begin{center}
   \includegraphics[width=7.5cm]{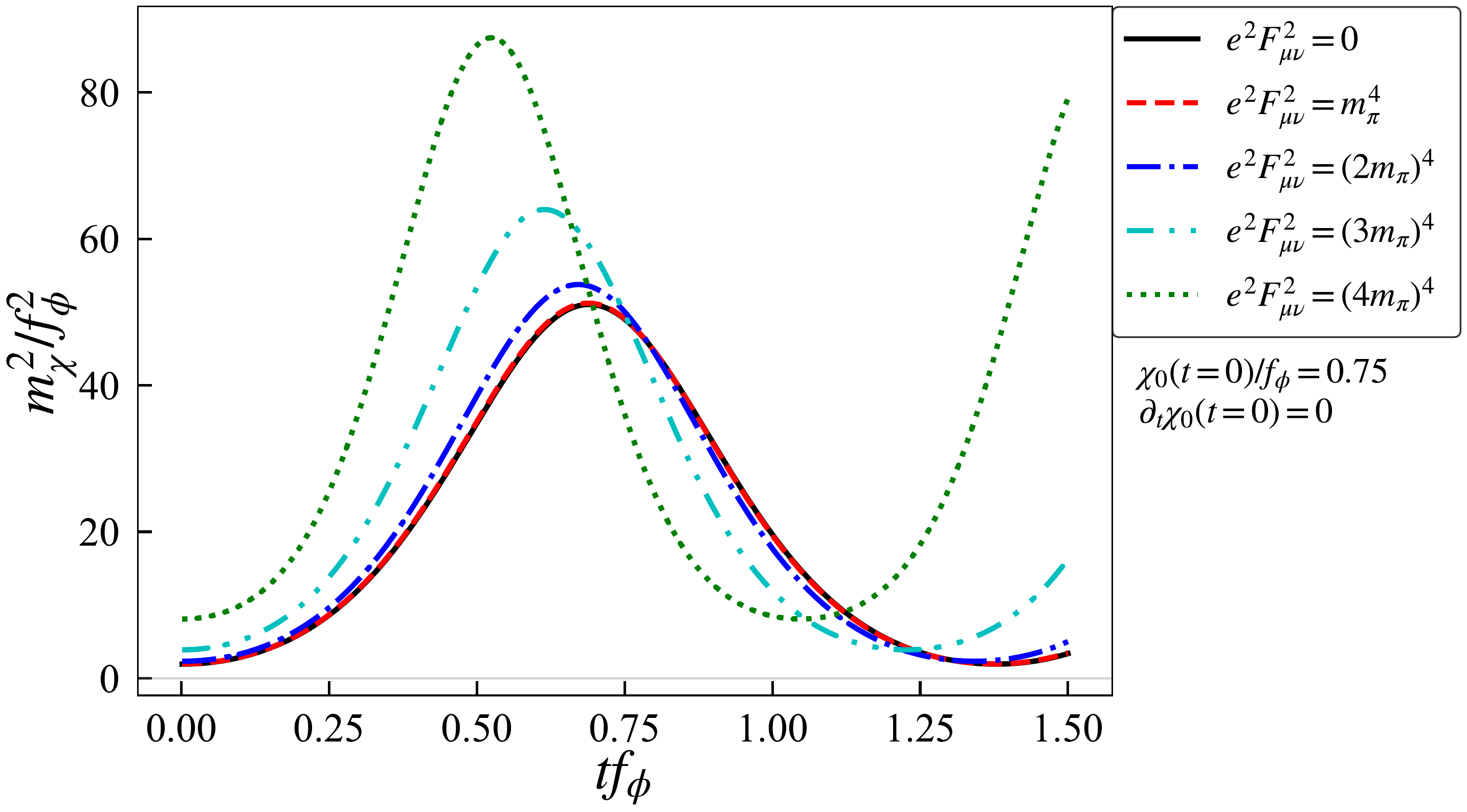}
    \subfigure{(b)}
  \end{center}
 \end{minipage}\\
 \begin{minipage}{0.5\hsize}
  \begin{center}
   \includegraphics[width=7.5cm]{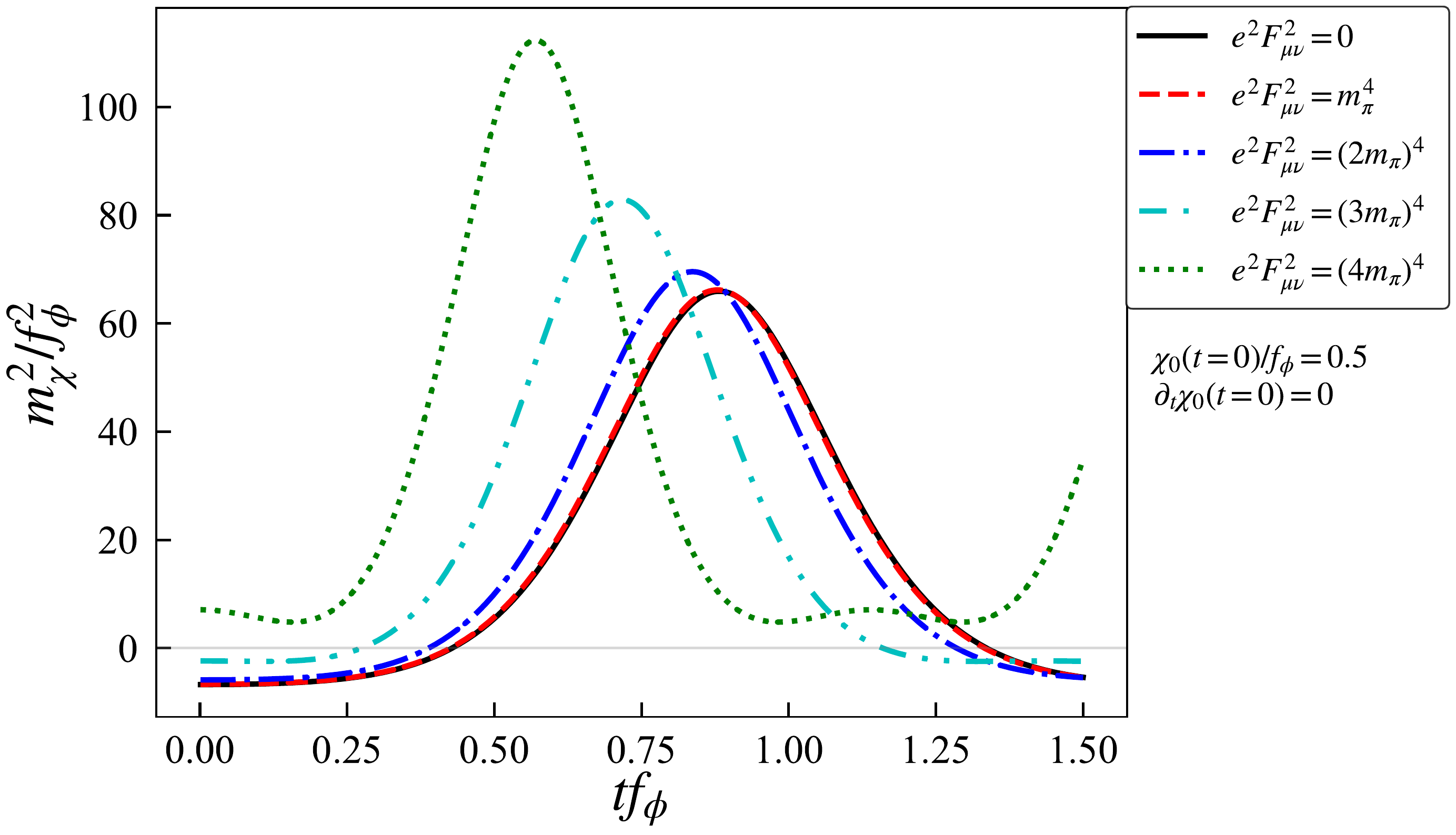}
    \subfigure{(c)}
  \end{center}
 \end{minipage}
 \end{tabular}
 \caption{ Time evolution of the effective ``mass" of the fluctuating dilaton
 in Eq.(\ref{mchi}),
for (a) $\chi_0(t=0)/f_\phi=1.0$, (b) $\chi_0(t=0)/f_\phi=0.75$ and (b) $\chi_0(t=0)/f_\phi=0.5$.
}
 \label{ploteffmass}
\end{figure}

So far, we have examined the background level for
the oscillatory dilaton physics and anomalous transports.
In this subsection, we turn to consider
the fluctuation of the dilaton under the oscillating background $\chi_0(t)$.
From the dilaton potential,
the time-dependent effective ``mass" for
the fluctuating dilaton is evaluated as
\footnote{
This $m_\chi (t)$ 
represents the frequency (square root of curvature) of the oscillation around the background $\chi_0$, and 
should not be confused with 
$m_\phi^{\rm eff}(D)$ in Eq.~(\ref{eff-mass}), 
which corresponds to the mass of the background dilaton at rest: $m_\phi^{\rm eff}(D) = \sqrt{1 + \frac{D}{2}} m_\phi$.  
}
\begin{eqnarray}
m_\chi^2(t)&=&\frac{\partial^2 V(\chi)}{\partial \chi^2 }\Biggl|_{\chi=\chi_0(t)}\nonumber\\
&=&
\chi_0^2(t)\frac{m_\phi^2}{f_\phi^2}\left[
3\log\frac{\chi_0(t)}{f_\phi}+1
\right]
+
\frac{1}{\chi_0^2(t)} \frac{\beta(e)}{2e}F_{\mu\nu}^2.
\label{mchi}
\end{eqnarray}
Note that the fluctuating dilaton rolls along the potential slopes as
depicted in Fig.~\ref{pot_cal}, and feels instantaneous curvatures
of the potential hills, which correspond to the effective ``mass" in Eq.(\ref{mchi}).
Therefore, the time evolution of the
effective ``mass" highly depends
on the initial condition for the oscillating background $\chi_0(t)$.
We plot the effective ``mass" in Fig.~\ref{ploteffmass},
varying the initial conditions,
such as $\chi_0(t=0)/f_\phi=1.0,\,0.75,\,0.5$ (the panels (a), (b) and (c),
respectively).

Note that even in the absence of the EM field,
the dilaton oscillates when we put the dilaton background on a steep hill,
far from the stationary point of the dilaton potential.
Actually, this
is reflected in the effective ``mass" in Eq.(\ref{mchi}),
as shown in Fig.~\ref{ploteffmass}.
When we put the dilaton background much far from the stationary point of the dilaton potential as an initial condition, the effective ``mass"
more intensely oscillates.
Of interest is that the fluctuating dilaton gets a negative mass squared, so that the dilaton (instantaneously) becomes tachyonic, as seen from the panel (c) of
Fig.~\ref{ploteffmass}.
Now, switching on the EM field, the effective ``mass" is amplified
by the screening effect, as was also observed in the static picture in Fig.~\ref{pot_cal}, and a strong enough EM field finally
prevents the dilaton from becoming tachyonic in a whole time-scale,
so the dilaton time evolution gets completely stabilized.
Actually this drastic screening effect will be significant
for the present system, i.e., the hadron phase
to keep in the thermal equilibrium,
as will be briefly discussed below.

\begin{figure}[t]
\begin{tabular}{cc}
 \begin{minipage}{0.5\hsize}
  \begin{center}
   \includegraphics[width=7.5cm]{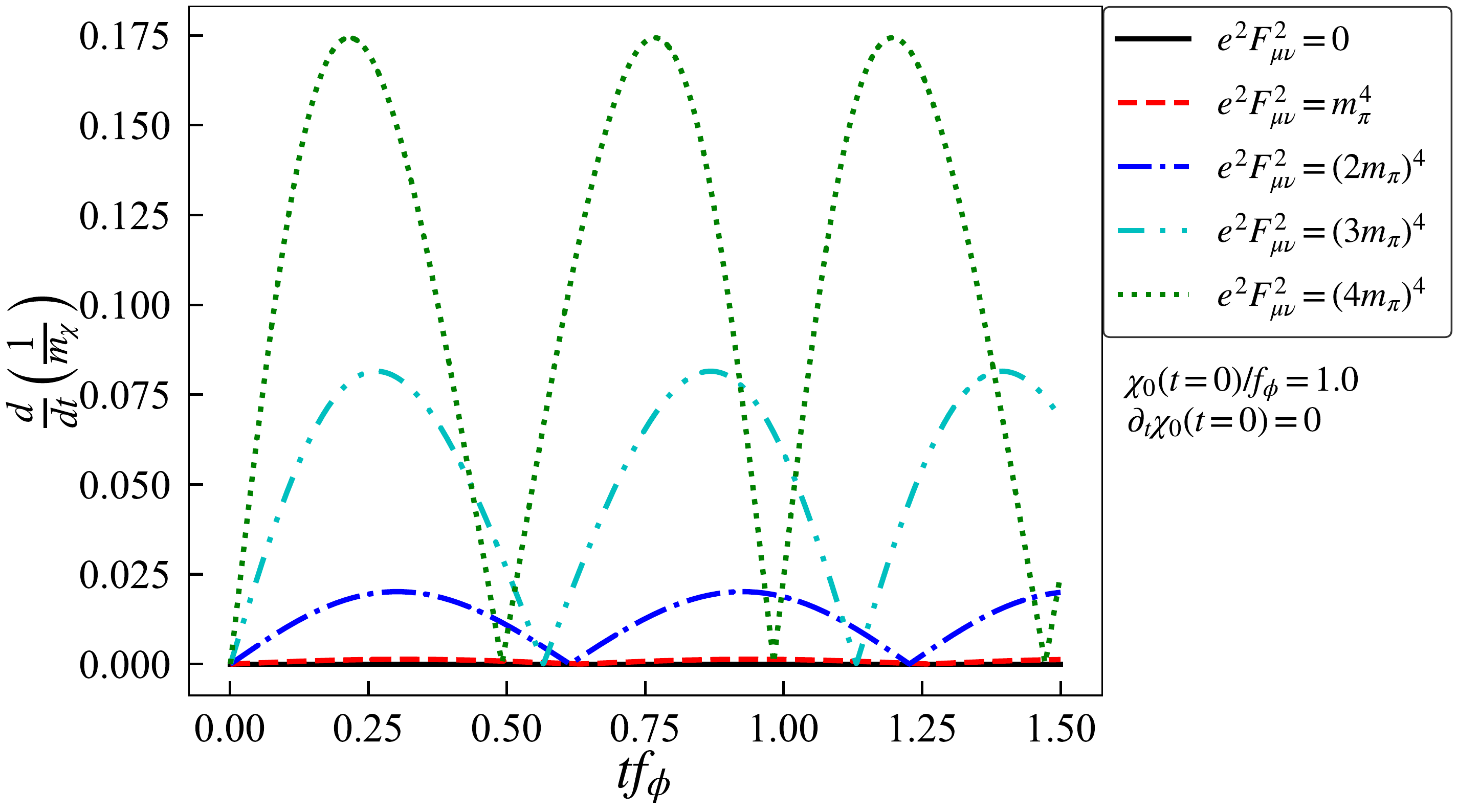}
    \subfigure{(a)}
  \end{center}
 \end{minipage}%
 \begin{minipage}{0.5\hsize}
  \begin{center}
   \includegraphics[width=7.5cm]{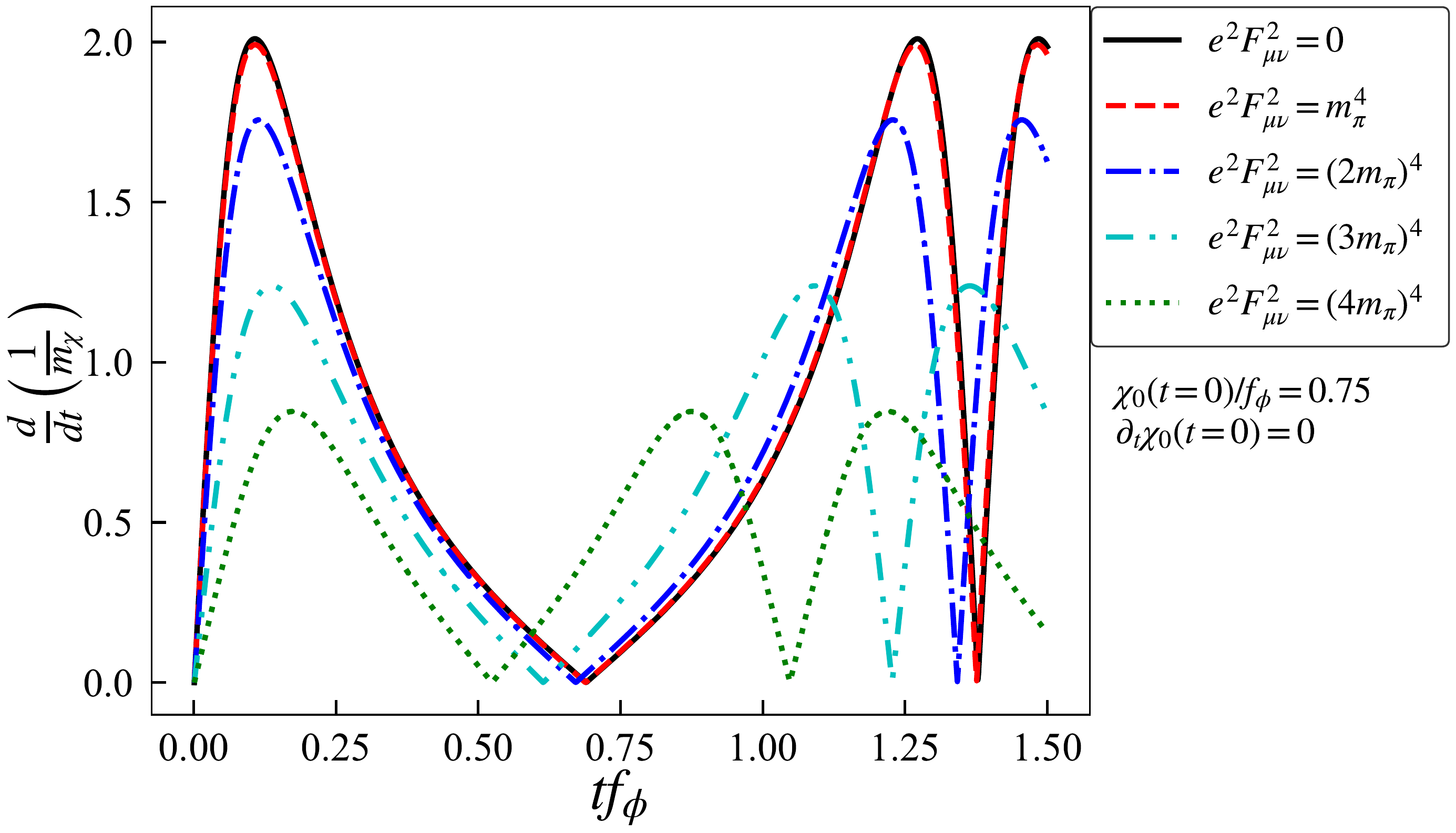}
    \subfigure{(b)}
  \end{center}
 \end{minipage}\\
 \begin{minipage}{0.5\hsize}
  \begin{center}
   \includegraphics[width=7.5cm]{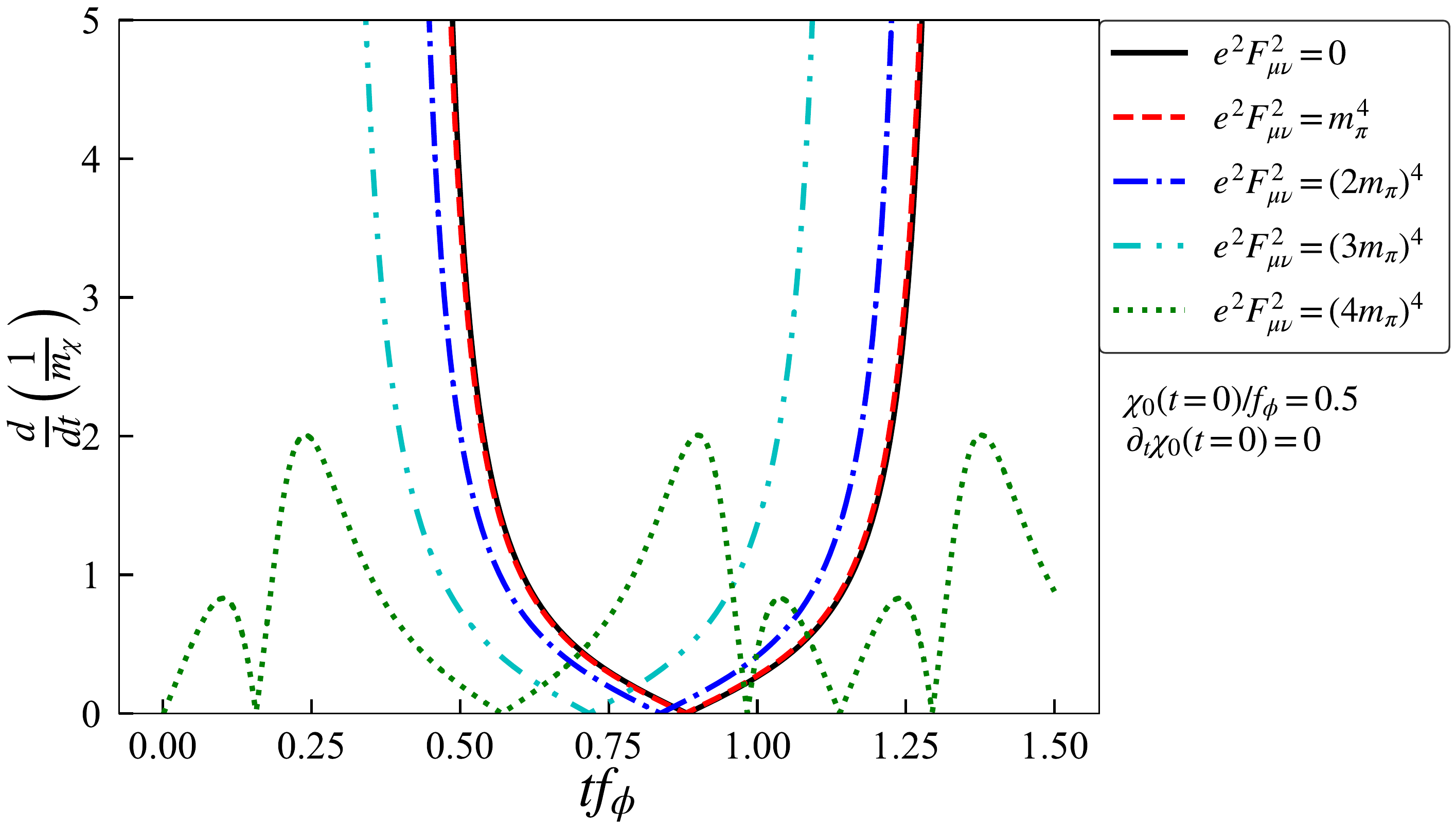}
    \subfigure{(c)}
  \end{center}
 \end{minipage}
 \end{tabular}
 \caption{ Plots of the nonadiabatic condition versus the normalized time scale
 $(t f_\phi)$, varying the initial conditions:
 (a) $\chi_0(t=0)/f_\phi=1.0$, (b) $\chi_0(t=0)/f_\phi=0.75$ and (c) $\chi_0(t=0)/f_\phi=0.5$.
}
 \label{non_ad}
\end{figure}

First, we should observe that
when the dilaton gets tachyonic,
the present target system, i.e., the hadron phase
is changed from the equilibrium (adiabatic)
to be out-of-equilibrium (nonadiabatic),
and then
nonperturbative particle productions would be generated by
the out-of equilibrium processes, to ``reheat"
the hadron phase by the explosively produced particles
(coupled to photons and photons themselves),
that is called the tachyonic preheating mechanism~\cite{Felder:2000hj}.
In that case,
the perturbative estimations for
the dilepton production in Eq.~(\ref{pro_diplepton}) and the diphoton production in Eq.~(\ref{pro_diphoton})
might be too naive, or significantly underestimated (or could be overestimated
due to highly nontrivial cancellations).
To check this point,
we evaluate the
non-adiabatic condition,
\begin{eqnarray}
\left|\frac{d}{dt}\left(\frac{1}{m_\chi (t)}\right)\right|
=
\left|\left(\frac{\dot m_\chi (t)}{m_\chi^2 (t)}\right)\right|
\gtrsim
1,
\end{eqnarray}
which is visualized in the plots depicted in Fig.~\ref{non_ad}.

The panel (a) shows that
at $\chi_0(t=0)/f_\phi=1.0$
the EM field acts as a catalyzer for the nonadiabaticity of the hadron phase system.
This catalysis can be interpreted as the ``kick" by the EM field,
to make the stationary dilaton highly active.
However, for weak EM fields satisfying $e^2 F_{\mu\nu}^2 \ll (4.0\, m_\pi)^4$,
the system cannot be nonadiabatic.
This implies that
the ``kicks" are still too small to highly accelerate
the stable dilaton.
Therefore, our perturbative
estimates on the anomalous productions in Eqs.(\ref{pro_diplepton})
and (\ref{pro_diphoton}), with such a weak EM field taken into account,
are reliable and intact.

For $\chi_0(t=0)/f_\phi=0.75$, the dilaton is already rolling
on the potential and oscillates.
As the EM field is turned on and increases,
the interference on the nonadiabaticity gets suppressed
as depicted in the panel (b) of Fig.~\ref{non_ad}.
Thus, for this initial condition,
we find that the EM field serves as an inverse catalysis (i.e. screening)
for the nonadiabaticity.

Finally, we examine the case where
the initial dilaton is put on a much higher hill in the potential
($\chi_0(t=0)/f_\phi=0.5$).
The panel (c) of Fig.~\ref{non_ad} shows that
 $|\frac{d}{dt}\left(1/m_\chi (t)\right)|$ diverges for
 $e^2 F_{\mu\nu}^2< (3m_\pi)^4$.
 Note that this is due to the EM effect, but
because the dilaton becomes tachyonic as was seen from the panel (c) of
Fig.~\ref{ploteffmass}.
When the EM field reaches a strong strength
 $F_{\mu\nu}^2= (4m_\pi)^4$,
the nonadiabaticity gets dismissed, and
$|\frac{d}{dt}\left(1/m_\chi (t)\right)|$ turns to keep finite values.
Thus, eventually even for this initial dilaton,
the EM field acts as the screening
effect on the nonadiabaticity.

To summarize, as long as the dilaton initially rolls on the potential, which correspond to
the cases with the initial conditions (b) and (c),
the EM field suppresses the nonadiabaticity to
keep driving the hadron phase system into the thermal equilibrium.
This is manifested by the screening effect caused by the EM field background,
and would be similar to the screening interference phenomenon
argued in Refs.~\cite{Kofman:1997yn,Czerwinska:2016mbm} in a context of backreactions generated
by the produced plasma in the preheating scenarios.


\section{Summary and discussion}

In this paper,
we have discussed a
scale anomalous current induced from QCD coupled to
an EM field, based on a dilaton effective theory
reflecting the QCD scale anomaly in a proper way.
We first clarified that as long as a QCD dilaton is
introduced as if it were a generic singlet scalar universally coupled to the target system,
the EM-induced scale anomalous transport
takes a universal form, given by
the spacetime dependence of a dilaton or a scale
factor in a curved spacetime (Eq.(\ref{ano_current})).
The form is completely fixed
by the scale/Weyl anomaly structure, whichever
one works on a curved or flat spacetime --
this was dubbed as an ``equivalent theorem",
followed from the frame/Weyl equivalence including
the anomaly.

We then claimed that
the QCD dilaton, being a dynamical particle contrast to non-dynamical scale factors,
would give a discrimination for
this robust universality by its dynamics.
Our target system has thus been chosen so as to have
a nontrivial dynamics of the QCD dilaton:
that is a dynamic oscillation
in spatially
homogeneous hadron phase (with a homogeneous EM field)
expected to be created
in thermal history of universe, or heavy ion collision
experiments.
It turned out that the dilaton starts to roll in the potential, and oscillates
by a ``kick" of the EM field, even if it used to stay at the vacuum.
We then observed that the scale anomalous current
arises along with the dynamic oscillating dilaton background coupled to
the homogeneous EM field, and oscillates as well.

As the phenomenological implication for, e.g., the
heavy ion collision experiments,
we paid our attention to the
anomalous dilepton and diphoton productions generated
from the dynamic scale anomalous transport, see Eqs.(\ref{pro_diplepton}) and (\ref{pro_diphoton}).
We observed that characteristic peak structures would be seen for those invariant mass distributions, at the effective dilaton mass. This kinematic feature
is surely intrinsic, and has never been seen in the ordinary QCD hadron physics.
Thus, those particle productions would be crucial signals  as indirect detection of the presence of the induced
dynamic scale anomalous transport
in heavy ion collision experiments.

We also investigated more details of the dynamics of the oscillatory dilaton background.
First, we found that in a homogeneous EM field background,
the dilaton potential is deformed to have a steeper and deeper well than the one without the EM filed, so that the dilaton field more promptly rolls down to the stationary point of the dilaton potential: the effective
dilaton ``mass" (i.e., frequency of the oscillation) gets larger and the stationary point (determining the dilaton decay constant) is shifted
to be larger (Fig.~\ref{pot_cal}).
Then, the time evolution of the oscillating dilaton background and the induced anomalous transport current were analyzed by referring to the potential deformation, where we observed
the magnitudes of the oscillations get
amplified by increasing the EM strength
(Figs.~\ref{Tcon_cal} and ~\ref{current_cal}).

The dynamic oscillatory dilaton background
also affects the fluctuating dilaton field
and can create an out-of-equilibrium state due to the
nonadiabatic oscillation, which can be realized when
the fluctuating dilaton (instantaneously) gets tachyonic,
leading to an explosive
nonperturbative particle production (called the tachyonic preheating).
We observed that the EM field contributes to the
tachyonic preheating as a screening effect,
so that the nonadiabaticity gets diluted by a
strong EM field (as long as the initial dilaton
can have a kinetic energy), see Figs.~\ref{ploteffmass} and \ref{non_ad}.

In the present work, we have assumed the presence of a dilatonic meson
(identified as
the lightest isosinglet scalar meson, $f_0(500)$).
Note, however, that even if we do not assume such a dialatonic scalar,
the form of the lightest isosinglet meson
coupling to EM fields
is robust, and
necessarily couples to the scale anomaly in QCD,
as long  as the meson is isosinglet.
For instance, if we started from an alternative potential form different from 
the dilatonic potential in Eq.~(\ref{dilaton_pot}),  
we could get a similar diphoton and/or dilepton signal, but different in magnitudes. 
Precise measurements on the diphoton and/or the dilepton productions might 
make it possible to 
probe whether the QCD dilaton picture is viable or not. 
Thus, the presence of the dynamic scale anomalous transport is a generic
consequence of QCD, although the ``equivalence theorem" may or
may not be applied to the form of the transport current.
So, the dynamic scale anomalous transport and/or the dynamic
feature of the QCD scalar oscillation might trace one slice of
the thermal history of universe, and/or that of the created circumstance
in heavy ion collision experiments where strong EM fields are expected to exist.

We have found several dynamic features associated with the presence of
the scale anomalous transport induced from QCD in EM field,
that would give keys to open a new avenue for the anomalous transport physics in heavy ion collisions, in parallel to
the dynamic particle physics that has been developed
in the inflation/preheating scenario.
\\

In closing,
we give some comments on possible issues left with us.
\\

(1) Incidentally, 
in a context different from the scale anomaly, 
it has been argued that 
the chiral anomaly induced by an EM field 
also drives similar time-oscillations for 
condensates (i.e. background fields) of 
the scalar- and pseudoscalar-quark bilinear (chiral partner) states,  
to take a spiral form~\cite{Hayata:2013sea}. 
This time-dependent spiral structure is associated with 
the dynamic oscillatory chiral currents in the temporal (time) direction induced by the EM field, 
and hence, called the temporal chiral spiral. 
The essential difference between the dilaton and the chiral-partner 
($\sigma(t)$ and $\pi(t)$) oscillatory backgrounds is that:  
the former $(\chi(t))$ can be viewed 
as the chiral singlet component of $\sigma(t)$ and $\pi(t)$; 
$\chi(t)=\sqrt{\sigma^2(t) + \pi^2(t)}$, 
so it keeps constant (which is like an invariant radial length) 
in time in the temporal chiral spiral because of the chiral invariance.  
Therefore it cannot be affected by any chiral transformation or anomaly  
as they act as a phase rotation, that is, 
the dilaton (radial) mode is driven to oscillate only by the scale transformation (anomaly).

Though being essentially discriminated in QCD 
with respect to the anomalous 
transport sources, both two oscillating condensates may share 
similar phenomenological applications even apart from  QCD. 
As addressed in Ref.~\cite{Hayata:2013sea}, in fact,  
the temporal chiral spiral can be applied 
to condensed matter systems such as carbon nanotubes~\cite{Bockrath:1999}, fractional quantum Hall edge states~\cite{Wen:1990} and one-dimensional cold-atom systems~\cite{Guan:2013}.
Given this fact, it would be expected that our observation of the oscillating dilaton background as well as the dynamic scale anomalous transport 
would also be applied to the condensed matter systems.
Thus, our findings would also be relevant to give a new insight of the condensed matter physics, which is involved in one of interesting future directions.

(2) The EM beta function can include a nontrivial EM field dependence
arising through the up and down quark loops, which would give
higher order corrections in powers of $(e^2 F_{\mu\nu}^2)$ to
the dilaton-photon coupling. In the present study,
we have simply neglected such higher order corrections,
by assuming a somewhat weak EM field case (compared to
the intrinsic QCD scale $\sim$ 1 GeV).
However, the higher order corrections might have significantly been
sizable in evaluating the dilepton and diphoton productions,
and nonadiabaticity for the dilaton oscillation.
If it is the case, those production rates would dramatically
be enhanced, and the dilaton oscillation would be nonadiabatic
even if is ``kicked" from the stationary point (corresponding to the
initial condition (a), in Fig.~\ref{non_ad}).
In particular, the correction to the latter case could
give a great impact on the thermal history of the hadron phase in early universe:
the out-of-equilibrium state is induced by the nonadiabatic dilaton oscillation,
hence nonperturbative particle productions (perhaps with an excessive
entropy production) might happen in the hadron phase. 
Heavy-ion collision experiments could have a chance to probe
such out-of-equilibrium state as well. To rigorously check this point,
we need to nonperturbatively compute the dilaton-photon-photon
triangle diagram involving the nonperturbative quark propagators.
Though this sort of nonperturbative computations has never been done,
our current study would motivate people to work on this issue.

(3) 
Regarding the nonperturbative particle production phenomena,
the lattice simulation approach 
have already started to implement    
the preheating mechanism inspired by the axion inflation~\cite{Cuissa:2018oiw}. 
In the axion system, photons can be explosively and nonperturbatively 
produced by the nonadiabatic axion oscillation via the interaction, $a F\tilde F$ with $a$ being the axion field, which has been observed 
on the lattice. 
Similarly, it could be plausible on lattice QCD in which 
a QCD dilaton coupled to diphoton, with the dilaton being assumed only 
composed of the chiral-singlet part of $u\bar{u} + d\bar{d}$, and measure the nonadiabatic diphoton production,  
through the interaction $\phi FF$. 
Such a lattice study will qualitatively and numerically clarify our prediction of the characteristic 
diphoton production signal as well as nonperturbatively 
proving the $\phi FF$ coupling, and hence, we would encourage lattice QCD communities to perform 
this kind of simulations in the future. 

(4) Another remark is on the stability of the oscillating dilaton background, i.e. the vacuum in the spatially
homogeneous hadron phase.
In the present study,
we have simply assumed the dilaton vacuum is completely
stable, so the dynamic oscillation keeps eternally.
However, in reality, the oscillating vacuum should have
the lifetime, and decay in a finite time scale
(presumably before the Big Bang nucleosynthesis for the 
thermal history of universe).
Or, the oscillation might be ended by a ``matter-plasma  effect" acting as a backreaction generated 
during nonperturbative particle productions~\cite{Dolgov:1989us,Kofman:1997yn,Traschen:1990sw,Kofman:1994rk}.

Inclusion of this lifetime can also be done by taking into account
possible imaginary parts in the Hamiltonian, which could arise from
quantum loop effects, or alternatively along the line of Ref.~\cite{Huang:2019phi} in which the decay of an oscillatory axion-condensed vacuum is studied.

(5) A final remark is about the homogeneity assumption we used in our calculation. Going beyond this assumption would lead to more interesting phenomenological implications. 
First of all, the possible inhomogeneity in heavy ion collisions 
would smear 
the kinematical peak structure predicted in the dilepton 
and diphoton productions (Eqs.(\ref{pro_diplepton}) and (\ref{pro_diphoton})), 
because 
the dilaton background would become spatially  
inhomogeneous to make the original resonance form (like a delta function) 
smeared. 
This smearing effect might be observed 
analogously to a collisional broadening in medium, 
and make the production amplitudes modestly smaller.

Second, if we consider an extreme case of inhomogeneity, namely, 
a ``boundary" in space, the scale anomaly was shown to induce a boundary current (and/or boundary charge density) in the presence of the EM field, as discussed in~\cite{Chu:2018ksb,Chu:2018ntx,Zheng:2019xeu,Chu:2020mwx,Chu:2020gwq,Chernodub:2019blw}. 
It would also be deserved to examine how the ``equivalence-theorem" for the scale anomalous transport works even in such an extremely inhomogeneous condition and how the boundary current are understood in the dilaton effective theory.

These issues would be important to give more precise
predictions to the anomalous dilepton and diphoton
productions, the nonadiabatic oscillatory
dilaton ``phase" as above, and also other related topics. 
Their interesting subjects are to be pursued
in the future.

\section*{Acknowledgements}

We are grateful to Kenji Fukushima for enlightening discussions and Seishi Enomoto for useful comments on the screening effect on the tachyonic preheating. 
We also thank Rod Crewther for useful comments on a QCD dilaton acting as a Nambu-Goldstone boson.
S.M. was supported in part by the National Science Foundation of China (NSFC) under Grant No.~11747308 and No.~11975108 and the Seeds Funding of Jilin University (S.M.). M.K. thanks for the hospitality of Center for Theoretical Physics and College of Physics, Jilin University where the present work has been partially done. X.G.H was supported in part by NSFC under Grant No.~11535012 and No.~11675041.



\end{document}